\documentclass[a4paper]{article}
\usepackage{fullpage}

\usepackage[doi=false,
isbn=false,
url=true,
sorting=nyt,
eprint=false,
maxbibnames=5,	
giveninits=true,
backend=bibtex,
style=alphabetic]{biblatex}		
\usepackage[normalem]{ulem} 

\usepackage{hyperref}		
\hypersetup{
    colorlinks=true, 
    linkcolor=blue, 
    citecolor=red, 
    urlcolor=red 
}

\usepackage{multicol}

\usepackage{amsfonts}		
\usepackage{amsmath}		
\usepackage{amssymb}	    
\usepackage{tikz-cd}		
\usepackage{amsthm}			
\usepackage{mathtools}      

\usepackage{bm}     

\usepackage{algorithm}     
\usepackage{algpseudocode}    

\usepackage{xspace}         
\usepackage{todonotes}      
\usepackage{mathabx}        
\usepackage[mathscr]{euscript}
\DeclareMathAlphabet{\mathpzc}{OT1}{pzc}{m}{it} 
\usepackage{amsmath,amssymb,mathtools}
\usepackage{nicematrix} 
\usepackage{enumitem}
\usepackage[cal=boondox]{mathalfa}
\usepackage{relsize}

\newcommand{\ms}{[m_s]}
\newcommand{\mt}{[m_t]}
\newcommand{\md}{[m_d]} 
\newcommand{\Ns}{N_s} 

\newcommand{\Z}{\mathbb{Z}}
\newcommand{\N}{\mathbb{N}}
\newcommand{\Q}{\mathbb{Q}}
\newcommand{\R}{\mathbb{R}}
\newcommand{\C}{\mathbb{C}}

\newcommand{\K}{\mathbb{K}}
\newcommand{\Sr}{\mathbb{S}}
\newcommand{\PP}{\mathbb{P}}

\newcommand{\VV}{\mathbb{V}}

\newcommand{\VVC}{\mathbb{V}_{\CC}}

\newcommand{\none}[1]{\lVert #1 \rVert_1}
\newcommand{\ntwo}[1]{\lVert #1 \rVert_2}
\newcommand{\ninf}[1]{\lVert #1 \rVert_{\infty}}

\newcommand{\X}{\bm{X}}

\newcommand{\x}{\bm{x}}

\newcommand{\z}{\bm{z}}

\newcommand{\vv}{\mathbf{v}}

\newcommand{\balpha}{\bm{\alpha}}

\newcommand{\diag}{\mathtt{diag}}

\newcommand{\Igrad}{I_{\nabla f}}

\newcommand{\Irur}{I_{\mathtt{RUR}(\nabla\! f)}}

\newcommand{\ideal}[1]{\langle #1 \rangle}

\newcommand{\GB}{\text{Gr\"{o}bner }}

\newcommand{\disc}{\mathbb{D}}

\usepackage{xparse}

\NewDocumentCommand{\wt}{m}{%
  \widetilde{#1}%
}

\DeclareDocumentCommand\Rad{ g }{%
    { \mathtt{R} \IfNoValueF {#1} {(#1)} } }

\DeclareDocumentCommand\nsteo{ g }{%
    { \mathscr{S} \IfNoValueF {#1} {(#1) } } }

\DeclareDocumentCommand\pospert{ g }{%
    { \mathscr{P}_{\lambda} \IfNoValueF {#1} {(#1) } } }   
\DeclareDocumentCommand\negpert{ g }{%
    { \mathscr{N}_{\lambda} \IfNoValueF {#1} {(#1) } } }       

\DeclareDocumentCommand\bd{ g }{%
    { \mathtt{B} \IfNoValueF {#1} {(#1) } } }

\newcommand{\pertval}{\mathscr{E}}		

\DeclareDocumentCommand\fe{ g }{%
    { f_{\pertval} \IfNoValueF {#1} {(#1) } } }

\newcommand{\eps}{\varepsilon}
\newcommand{\feps}{f_{\eps}}
\newcommand{\fte}{f_{t,\eps}}

\newcommand{\bsz}{\ensuremath{\mathlarger{\mathcal{h}}}\xspace}

\newcommand{\abs}[1]{| #1 |\xspace}

\newcommand{\NN}{\ensuremath{\mathbb{N}}\xspace}
\newcommand{\ZZ}{\ensuremath{\mathbb{Z}}\xspace}
\newcommand{\QQ}{\ensuremath{\mathbb{Q}}\xspace}
\newcommand{\RR}{\ensuremath{\mathbb{R}}\xspace}
\newcommand{\CC}{\ensuremath{\mathbb{C}}\xspace}

\newcommand{\OO}{\ensuremath{\mathcal{O}}\xspace}
\newcommand{\sO}{\ensuremath{\widetilde{\mathcal{O}}}\xspace}
\newcommand{\OB}{\ensuremath{\mathcal{O}_B}\xspace}
\newcommand{\sOB}{\widetilde{\mathcal{O}}_B\xspace}

\newcommand{\rur}{\mathtt{RUR}\xspace}

\newcommand{\sosrur}{{\rm\texttt{SOS-RUR}}\xspace}

\newcommand{\hjpos}{{\rm\texttt{HJ-SOS-POS}}\xspace}
\newcommand{\hjneg}{{\rm\texttt{HJ-SOS-NEG}}\xspace}
\newcommand{\hjsosrur}{{\rm \texttt{HJ-SOS-RUR}}\xspace}

\newtheorem{thm}{Theorem}[section]
\newtheorem{cor}[thm]{Corollary}
\newtheorem{prop}[thm]{Proposition}
\newtheorem{lem}[thm]{Lemma}

\newtheorem{dfn}[thm]{Definition}
\newtheorem{examp}[thm]{Example}

\newtheorem{rmk}[thm]{Remark}

\title{Certificates for nonnegativity of multivariate integer polynomials under perturbations}
\author{
Mat\'ias Bender~\thanks{Inria Saclay and CMAP, \'Ecole Polytechnique, IP Paris, France}
\and Khazhgali Kozhasov
\thanks{Universit\'e C\^ote d'Azur, France}
 \and
Elias Tsigaridas~%
\thanks{Inria Paris and IMJ-PRG, Sorbonne Universit\'e, France}
\and
Chaoping Zhu~%
\thanks{Inria Paris and IMJ-PRG, Sorbonne Universit\'e, France}
}

\begin{document}

\maketitle

\begin{abstract}

We develop a general and unconditional framework 
for certifying the global nonnegativity of multivariate integer polynomials; based on rewriting them as sum of squares modulo their gradient ideals.
We remove the two structural assumptions typically required by
other approaches, namely that the polynomial attains its infimum 
and zero-dimensionality of the gradient ideal.
Our approach combines a denominator-free stereographic transformation 
with a refined variant of the Hanzon–Jibetean perturbation scheme. 
The stereographic transformation preserves nonnegativity 
while making the polynomial coercive, with explicit bounds 
on the radius of positivity and on the nonzero critical values.
Subsequently, we apply carefully constructed explicit perturbations 
that enforce zero-dimensionality of the gradient ideal without 
altering nonnegativity, allowing us to invoke recent algorithms 
to derive algebraic certificates or rational witness points.
 We present three algorithms implementing our framework 
 and analyze their bit complexity in detail, 
 which is single exponential with respect to the number of variables.

A second contribution is a new explicit SOS perturbation scheme, which
allows us to perturb any {nonnegative} polynomial in such a way that
it can be written as a sum of squares (SOS).
In contrast to Lasserre’s classical SOS approximation, 
which guaranties density but currently does not provide an effective control 
over the perturbation size, we only derive concrete perturbation bounds ensuring that {a nonnegative polynomial enters the SOS cone.}

\end{abstract}

\medskip
\textbf{Keywords:}
	Nonnegative polynomials, certificates of positivity,
	stereographic projection, sum of squares decomposition, sum of squares approximations,
	gradient ideal, bit complexity

\medskip
\textbf{AMS subject classifications:}
		14Q30, 
        14N05, 
        11E25, 
        13P15, 
        90C23 

\newpage    

\tableofcontents

\newpage

\section{Introduction}

A central topic at the intersection of real algebraic geometry and polynomial optimization is
to decide whether a multivariate polynomial of even degree is nonnegative over $\RR^n$.
Deciding nonnegativity also allows us to verify whether a given value is a global lower bound for a multivariate polynomial; this simple observation has culminated in the development of a vibrant area of optimization. 
As this decision problem is $\text{NP}$-hard already for polynomials of degree $4$ \cite{MurtyKabadi1987},
current algorithms are computationally demanding.
Hence, it naturally arises the question of verifying their output, namely, nonnegativity, without having to solve
the optimization problem itself; this is what we call a
\emph{certificate of nonnegativity}.

For a polynomial nonnegative on $\RR^n$, the ultimate certificate is Artin's solution \cite{artin_uber_1927} to Hilbert's $17$th problem,
which ensures the existence of a representation of any nonnegative polynomial as a sum of squares of rational functions.
Unfortunately, the state-of-the-art degree bounds for the numerators and denominators in such representations are
prohibitive \cite{lpr-H17-20}. 
This motivates the search for certificates of nonnegativity that are easier and less expensive to compute in important special cases.

Following Powers \cite{powers-certif-book}, a certificate consists of one or more ``simple'' algebraic identities
which, if satisfied, imply the nonnegativity of the polynomial under consideration.
Such a certificate must be \emph{perfectly complete}, it should always certify a nonnegative polynomial, and \emph{perfectly sound}, it should correctly reject any polynomial that is not nonnegative; see \cite{btz-sosrur-hal-25} for further details.

We consider global certificates of nonnegativity for a polynomial $f \in \ZZ[\X] := \ZZ[X_1,\dots,X_n]$.
Given $f$, our aim is to provide algebraic identities that verify that $f(\bm{x}) \ge 0$ for all $\bm{x} \in \RR^n$,
or, if this is not the case, a point $\bm{p} \in \QQ^n$ such that $f(\bm{p}) < 0$, which we call a \emph{witness} point.
Existing approaches, discussed in more detail in the sequel, are powerful but typically require structural assumptions on $f$
(e.g., that the gradient ideal of $f$ is zero-dimensional and/or radical, or that the infimum of $f$ over $\RR^n$ is attained).
However, many classical examples violate these assumptions, and it is of clear interest to develop alternative certificates of nonnegativity without imposing any conditions on the input polynomial.

\medskip
\begin{center}
We show that, by performing suitable invertible transformations, one can transform any nonnegative polynomial into another one whose gradient ideal is zero-dimensional and whose infimum is attained.
In this way, we can leverage existing methods for zero-dimensional systems (see for example \cite{btz-sosrur-hal-25}),
to provide global certificates of nonnegativity for an arbitrary input polynomial {with integer coefficients}. 
\end{center}
\medskip

\paragraph{Prior works}

The most commonly used certificate of nonnegativity is the SOS certificate, that is,
decomposing $f$ as a sum of squares of polynomials or rational functions.
In the univariate case, it is well known that any $f \in \RR[X]$ that is nonnegative can be written as a sum of two squares in $\RR[X]$~\cite{Hilbert1888}.
Over the rational numbers, the strongest general result is that if $f \in \QQ[X]$ is nonnegative, then it admits a weighted sum of squares representation with $t \le 5$ squares,
	$f(X) = \sum_{i=1}^t w_i\, s_i^2(X)$,
where $s_i \in \QQ[X]$, $w_i \in \QQ_+$, 
and $t \leq 5$~\cite{landau_uber_1906, pourchet_sur_1971}.
Unfortunately, no known algorithm achieves such a decomposition with $t \leq 5$, although there has been recent significant progress~\cite{kmv-prouchet-23}; see also~\cite{bdafmm-eff-pourchet-25}.
By allowing up to $t \leq \deg f + 3$ summands, the algorithms of~\cite{schweighofer_algorithmische_1999, chevillard_efficient_2011} provide constructive decompositions, with complexity analyzed in~\cite{mss-usos-19}; see~\cite{bdt-unisos-25} for the most recent improvements.

For multivariate polynomials, certificates of nonnegativity are considerably more intricate.
The first major mathematical breakthrough is Artin’s solution to Hilbert’s 17th problem~\cite{artin_uber_1927}, which establishes that every real nonnegative polynomial is a sum of squares of rational functions.
Although Lombardi, Perrucci, and Roy~\cite{lpr-H17-20} give an algorithmic realization of Artin’s theorem, its complexity is prohibitive, in particular, the degree bounds for the numerators and denominators form a tower of five exponentials, rendering the method impractical.
Unlike the univariate case, not all multivariate nonnegative polynomials are sums of squares, with Motzkin’s polynomial
$M(\X) = X_1^4X_2^2 + X_1^2X_2^4 - 3X_1^2X_2^2 + 1$
providing the first explicit example.
Scheiderer’s striking result~\cite{scheiderer_sums_2016} further shows that there exist nonnegative polynomials in $\Q[\X]$ that are SOS in $\R[\X]$ but not in $\Q[\X]$.
Semidefinite programming hierarchies~\cite{lasserre_global_2001, parrilo_structured_2000} introduced weighted SOS relaxations that approximate nonnegativity; these methods are particularly effective in optimization due to the availability of numerical solvers.
Alternative certificates include SAGE/SONC polynomials~\cite{2014RelativeER, Magron-Seidler-Wolff-Circuits-2008, Wang-Magron-circuits-2020}, though they suffer from analogous limitations, as general nonnegative polynomials need not admit SAGE/SONC representations.

Because of these limitations, SOS representations cannot serve as universal, computationally affordable certificates of nonnegativity.
However, when the input polynomial satisfies certain structural assumptions (such as radicality and/or zero-dimensionality of the gradient ideal, or attainment of the global minimum), one can express $f$ as a sum of squares modulo its gradient ideal.
The \emph{gradient ideal} of $f \in \RR[\X]$ is {defined as the ideal generated by its partial derivatives,} 
\[
  \Igrad \coloneq \ideal{\partial_1 f, \dots, \partial_n f} \subseteq \RR[\X],
\]
and the associated \emph{gradient variety} $\VV(\Igrad) \subseteq \CC^n$ consists of all (complex) critical points of $f$.
First, 
Lasserre~\cite{lasserre-sos-grid-02}  gave an SOS certificate for nonnegativity of polynomials over finite Cartesian grids in $\RR^n$, and then, Parrilo~\cite{Parrilo-explicit-sos-certif-tr} obtained a more general certificate over finite point sets defined by radical ideals.
Nie, Demmel, and Sturmfels~\cite{nie_minimizing_2006} proved that $f$ is nonnegative on the real locus $\VV(\Igrad)\cap \RR^n$ of its gradient variety if and only if $f$ is a sum of squares modulo the radical of the gradient ideal $\sqrt{\Igrad}$.
Assuming that $f$ attains its infimum, this yields the certificate
\[
  f \ge 0 \text{ on } \RR^n \iff f = \mathrm{SOS} \bmod \Igrad,
\]
provided $\Igrad$ is radical.

Let $f \in \QQ[\X]$ have degree $d$ and maximum coefficient bitsize
$\tau$.  Magron, Safey El Din, and Vu~\cite{magron_sum_2023} gave an
algorithm that partially realizes the above certificate under the
assumptions that the gradient ideal $\Igrad$ is zero-dimensional,
radical and that the $X_1$-coordinates of the critical points are
pairwise distinct.  The algorithm first computes a representation
\[
  \VV(\Igrad)
  = \left\{
      \x \in \CC^n \,:\,
      w(x_1)=0,\;
      x_2=\frac{\kappa_2(x_1)}{w'(x_1)},\dots,
      x_n=\frac{\kappa_n(x_1)}{w'(x_1)}
    \right\},
\]
where $w, \kappa_2,\dots,\kappa_n \in \Z[X_1]$ are univariate polynomials and $w'$ is the derivative of $w$.
Then $f$ is nonnegative if and only if it admits a decomposition
\begin{equation*}
  \label{eq:f-r-sos-old}
  f
  = \frac{1}{(w')^d}
    \sum\nolimits_{j=1}^s c_j\, q_j^{2}
    + \sum\nolimits_{i=2}^{n}
        \left(X_i - \frac{\kappa_i}{w'}\right)\phi_i,
\end{equation*}
where $s \in \ZZ$, $q_j \in \ZZ[X_1]$, $\phi_i \in \ZZ[\X]$, and $c_j \in \QQ_+$ for all $i \in [n]$ and $j \in [s]$.
In particular, $\sum_{j=1}^s c_j q_j^2 \in \ZZ[X_1]$ is the remainder of $(w')^d f$ upon division by the divisors
$\{w'X_2 - \kappa_2, \dots, w'X_n - \kappa_n\}$.
When $f$ is nonnegative, this remainder is a nonnegative univariate polynomial and thus a weighted sum of squares in $\Z[X_1]$~\cite{chevillard_efficient_2011, mss-usos-19}.

More recently, Baldi, Krick, and Mourrain~\cite{bkm-posnullQ-arxiv} provided an SOS certificate of $f$ over a finite semi-algebraic set under specific assumptions on its defining polynomials.
Furthermore,~\cite{btz-sosrur-hal-25} obtained a similar certificate of nonnegativity based on the rational univariate representation of the roots, valid when the infimum of $f$ is attained and the gradient ideal is zero-dimensional; in particular, this removes the radicality assumption.
This yields the \sosrur algorithm, which we also employ in our setting; see Sec.~\ref{sec:sos-mod-grad-ideal} for further details.

When the infimum of $f$ over $\RR^n$ is not attained, instead of considering the gradient variety of $f$ and its values on it, one can look at \emph{asymptotic critical values}. A complex number $c\in \RR$ is called an asymptotic critical value of $f$, if there exists a sequence $(x_k)_{k\in \NN}\subset \CC^n$ tending to infinity, that is, $\Vert x_k\Vert\xrightarrow{k\rightarrow \infty} +\infty$, such that $f(x_k)\xrightarrow{k\rightarrow\infty} c$ and $\Vert x_k\Vert\cdot\Vert \nabla f(x_k)\Vert \xrightarrow{k\rightarrow\infty} 0$. There are algorithms to compute asymptotic critical values relying on Gröbner bases, e.g., \cite{KurdykaOrroSimon2000, JelonekKurdyka2003, SafeyElDin2007},
	or resultants~\cite{elt-inf-arxiv-24}.
However, though these algorithms allow us to compute these asymptotic values, it is not clear how to transform them into certificates of nonnegativity.
An important work on this direction is due to Schweighofer
\cite{Schweighofer2006} who powered the classical SOS approach to global minimization of a polynomial with the idea to certify nonnegativity over so called gradient tentacles, which are sets stretching out to infinity and that contain asymptotic minimizing sequences.

In general, perturbation methods are used to transform a possibly degenerate problem
(for example, one with solutions of high multiplicity, non-proper intersections,
non-surjective projections, etc.) into one with generic properties.
In the context of polynomial optimization, we may regard as degenerate those
configurations in which the gradient ideal is not zero-dimensional or the polynomial
does not attain its infimum, to mention a few important ones.
Parrilo and Sturmfels~\cite{pablo_a_parrilo_bernd_sturmfels_minimizing_2003}
considered this problem for polynomials having a finite number of critical points.
They computed their minimum by determining all critical points (explicitly or implicitly)
and then checking the sign of the finite set of corresponding critical values.
These critical points may be obtained by Gröbner bases, resultants, or homotopy continuation techniques.
Hanzon and Jibetean~\cite{hanzon2003global} provided a symbolic method for the global
minimization of multivariate polynomials, without assuming convexity or coercivity.
Given a polynomial $f$, they introduced the parametrized family of perturbations
\[
  f_{\lambda} := f + \lambda (X_1^{2m} + \dots + X_n^{2m}),
\]
where $2m > \deg f$.
For any fixed positive $\lambda$, these perturbed polynomials have finitely many
critical points and their minimum is always attained.
As the perturbation parameter tends to zero, the method recovers the infimum of the
original polynomial.
Thus, this approach provides a unified treatment of the cases where a minimum exists,
where only a finite infimum exists, and where the infimum is attained “at infinity”.
The main drawback is that one is forced to work with polynomials whose coefficients
are themselves polynomials in~$\lambda$.
Later, Jibetean and Laurent~\cite{jibetean2005semidefinite} combined this perturbation
scheme with Lasserre’s SOS hierarchy to provide certificates for the infimum using
sums of squares.
Our approach builds on Hanzon and Jibetean’s perturbation framework,
but we specialize $\lambda$ to avoid working with polynomials having other polynomials as coefficients.

The geometric viewpoint of polynomial optimization also provides a wealth of information.
The cones of nonnegative and SOS polynomials are fundamental in this setting.
However, on the one hand, deciding the positivity of a polynomial is a
difficult problem, and the boundary structure of the nonnegativity
cone is highly intricate. 
On the other hand, verifying whether a polynomial admits an SOS representation reduces to checking the positive semidefiniteness of an associated Gram matrix~\cite{powers-certif-book, clr-real-sos-95}, and therefore to solving a semidefinite feasibility problem.
Hence, the SOS cone is a (rather) tractable inner approximation of the
nonnegativity cone, and a substantial body of work studies how well
SOS polynomials approximate globally nonnegative ones. 

One can quantify the discrepancy between the two cones in several ways.
For fixed degree, Blekherman~\cite{blekherman_there_2006} showed that the volume of the nonnegativity cone is overwhelmingly larger than that of the SOS cone, and that the ratio grows superpolynomially with the number of variables; see also~\cite{ergur-mhomo-sos-18} for a similar result for multihomogeneous polynomials.
At the opposite end of the spectrum, when the degree is allowed to increase, a classical result of Berg~\cite{berg-mmom-87} shows that sums of squares are \emph{dense} in the cone of polynomials nonnegative on the hypercube $[-1,1]^n$ with respect to the $\ell_1$-norm of the coefficient vector.
Later, Lasserre and Netzer~\cite{l-sos-approx-06, ln-sos-pert-07} introduced explicit perturbation schemes that yielded fully explicit sequences of SOS polynomials converging to any given nonnegative polynomial in the $\ell_1$-topology; thus they quantified, in a sense, the density of SOS polynomials in the cone of nonnegative ones.
However, even though Lasserre's SOS approximation~\cite{l-sos-approx-06} leads to certificates of nonnegativity for the perturbed polynomial, it does not imply nonnegativity of the original polynomial.
The difficulty lies in the rate of convergence of the corresponding sequence: even when a perturbed polynomial is nonnegative, it is not known how small the perturbation must be in order to conclude that the original polynomial is nonnegative.
We introduce a new perturbation scheme, with explicit bounds on the perturbation, that places a nonnegative polynomial inside the SOS cone; unfortunately, this does not yet come with a convergence guarantee.


\paragraph*{Our contributions}

We introduce a general and unconditional framework to certify 
the global nonnegativity of a multivariate integer polynomial $f \in \ZZ[\X]$, 
based on certificates of positivity modulo their gradient ideal. 
We remove the two strong assumptions usually imposed by other approaches;
namely:
(i)~that the polynomial attains its infimum, and
(ii)~that its gradient ideal is zero-dimensional.
We no longer impose either assumption by relying on perturbations
to decide nonnegativity for any polynomial $f \in \ZZ[\X]$,
with $f(\bm{0}) >0$, without appealing to asymptotic critical values or semidefinite approximations.
For this, we employ a series of transformations
that preserve nonnegativity and enforce both conditions (i) and (ii); 
this allows to use \sosrur algorithm \cite{btz-sosrur-hal-25} to certify 
our input polynomial, see Def.~\ref{def:certPos} and Thm.~\ref{thm:SOS-RUR}.

First, we employ a denominator-free variant of stereographic transformation (Def.~\ref{dfn:steo-trans}).
 The transformed polynomial, $\nsteo{f}$, preserves the nonnegativity of $f$ (Thm.\ref{thm:aff-steo-equiv})
 and is coersive (Def.~\ref{def:coerciveness}). Even more, it holds (Thm.~\ref{thm:coercive-direct-steo}), for any $\x \in \RR^n$, 
 with $\ntwo{\x}^2 \geq 1$, that 
\begin{align}\label{eq:intro_eq}
\nsteo{f}(\x) \geq  \ntwo{\x}^{2d} - C \ntwo{\x}^{2d-1},
\end{align}
for a constant $C$ depending only on $(n,d, \bsz(f))$;
hence, we derive an explicit radius outside which $\nsteo{f}$ is strictly positive (Cor.~\ref{cor:steo-radius-estimates}).
Using a quantitative lower bound on the nonzero critical values, by Brownawell and Yap \cite{lower_bounds_Brownawell_Yap_2009} (Cor.~\ref{cor:boundCritValSteo}),
we form the backbone of our geometric reduction, as the behavior of $f$ at infinity can be normalized into a coercive regime through a transformation (with fully explicit complexity and bitsize estimates).
A second key property of $\nsteo{f}$ is that negative evaluations of $\nsteo{f}$ can be transported back to negative evaluations of $f$ (Lem.~\ref{lem:stereo-witness}).
If $\nsteo{f}(\x_0) < 0$ for some rational point $\x_0$, then we can explicitly construct a rational point $\mathscr{Q}(\x_0)$ such that $f(\mathscr{Q}(\x_0))<0$.  
Therefore, deciding the nonnegativity of~$f$ is algorithmically equivalent to deciding the nonnegativity of $\nsteo{f}$.

\medskip
\noindent
\emph{Hanzon--Jibetean-like perturbations and zero-dimensionality.}
Even though $\nsteo{f}$ is coercive and attains its infimum, its gradient ideal may still be positive-dimensional.
We enforce zero-dimensionality through a refined Hanzon--Jibetean perturbation~\cite{hanzon2003global}.  
We introduce positive and negative perturbations, $\pospert{f}$ and $\negpert{f}$
(Def.~\ref{def:pert-lambda}) {for a polynomial $f$ of degree $d$}, that is,
\[
\pospert{f}(\X) := f(\X) 
+ \lambda \sum\nolimits_{i=1}^{n} \left(1+X_i^2 + X_{i}^{d+2} \right) 
\quad \text{ and } \quad  
	\negpert{f}(\X) := f(\X) 
	- \lambda \sum\nolimits_{i=1}^{n}(1+X_i^2 + X_{i}^{d}) .
\] 
For small enough $\lambda$, both perturbations preserve nonnegativity. 
Even more, for any fixed $\lambda>0$ the gradient ideal of $\pospert{\nsteo{f}}$ is zero-dimensional and has no solutions ``at infinity''
(Thm.~\ref{thm:0-dim-plus-pert}). 
We then combine the coercivity and critical value bounds for $\nsteo{f}$ to obtain  explicit inequalities on~$\lambda$ 
(Thm.~\ref{thm:f-pos-effect-equiv-pert-steo})
 that ensure the preservation of nonnegativity. In particular, it holds 
\[
f \ge 0 
\quad\Longleftrightarrow\quad 
\pospert{\nsteo{f}} \ge 0.
\]
By combining these results with \sosrur (Thm.~\ref{thm:SOS-RUR}), we introduce the algorithm 
\hjpos (Alg.~\ref{alg:SOS-CS-POS})
that certifies the {nonnegativity of any input polynomial $f$ or computes 
a point at which it is negative}.

The main disadvantage of \hjpos is that it requires us to perform a very small
perturbations and it forces us, when the input consists of an integer polynomial,
to work with rationals having the worst case bitsize, right from the beginning. 
Even though this is not an obstacle from the theoretical perspective,
it limits the practicality of the algorithm.
To overcome this obstable, we exploit the negative perturbation, $\negpert{\nsteo{f}}$,
that allows us to gradually decrease the perturbations (Alg.~\ref{alg:SOS-CS-NEG}).
Unfortunately, this perturbation preserves nonnegativity
only when both $f$ and its highest degree homogeneous part are positive (Thm.~\ref{thm:minus-sign-perturbation-effective});
also, it does not provide a witness point, if $f$ is not nonnegative.
Nevertheless, it leads to the algorithm \hjneg (Alg.~\ref{alg:SOS-CS-NEG}) that certifies nonnegativity, under the conditions that we mentioned previously, 
and fails otherwise.

However, the combination of both perturbation schemes, and hence the combination
of the corresponding algorithms,
leads to the algorithm \hjsosrur (Alg.~\ref{alg:hjsosrur}) that incorporates 
the advantages of both approaches
and certifies the nonnegativity of any polynomial $f$,
without requiring us to work with small perturbations right from the beginning,
or demanding the nonnegativity of the highest degree homogeneous part. 

We study in detail the bit complexity of both \hjpos (Thm.~\ref{thm:sos-cs-pos-complexity}) and \hjneg (Thm.~\ref{thm:sos-cs-neg-complexity}),
as well as the bit complexity of \hjsosrur (Thm.~\ref{thm:sos-cs-complexity}).
We obtain single exponential, with respect to the number of variables, 
bit complexity bounds. In particular, if the input is a polynomial with $n$ variables, of degree $d$, and maximum coefficient bitsize $\tau$, then 
\hjsosrur in
\[	
	\sOB( e^{2 (\omega +1)n} \, (d+1)^{(\omega +2)n}\,  \tau)) 
\]
  bit operations, where $\omega$ is the exponent of matrix multiplication, 
  provides a certificate for $f$, involving rationals of bitsize at most 
  $\sO(2^{4n} (d+1)^{4n+2} \, ( d + \tau))))$.

 \medskip
 \noindent
 \emph{SOS under perturbation.}
  Finally, we consider the problem of expressing a positive polynomial as a sum of squares.
  As this is not possible in general, we consider a suitable perturbation
  to place it in the cone of SOS polynomials. 
  In particular, if $f$ is a positive polynomial in $n$ variables, of degree $2d$,
  then we show (Thm.~\ref{thm:sos-under-pert}) that for any $t \in \NN$, such that
	\[
	t  \geq \max \left\{ d,  
	\left\lceil 
	\frac{1}{\eps} \left( \none{f} +  \frac{\ntwo{\nabla{f}(\bm{0})}^2 }{f_0+ \eps} \right) 
	\right\rceil 
	\right\}, 
	\]
the degree $2t$ polynomial $\fte(\X) = f(\X)+\eps (1+\|\X\|_2^2)^t$ is a sum of squares.
This leads to the claim that, if $f$ is any positive polynomial in $n$ variables, of degree $2d$, then the polynomial $f(\X) + \eps (1 + \ntwo{\X}^2)^t$ is a sum of squares, if 
	\[
		t \geq \binom{n + 2d}{n} \, \frac{2^{\tau + 1}}{\eps}.
	\]
	This builds on the results of Lasserre~\cite{l-sos-approx-06} and Lasserre and 
	Netzer~\cite{ln-sos-pert-07}, but in our case we have explicit bounds
	on the power of the perturbation. Alas, our perturbation scheme
	does not have the property that $\fte \rightarrow f$, when $\eps \rightarrow 0$, cf.~\cite{l-norm1-approx-10,l-sos-approx-06}.

\paragraph{Organization}
The next section (Sec.~\ref{sec:preliminaries}) presents the preliminaries
from computational (real) algebraic geometry.
Next, in Sec.~\ref{sec:steo-transform} we present the stereographic transformation
and in Sec.~\ref{sec:HJ-pert} we present 
the Hanzon-Jibetean perturbations we exploit.
We present the algorithm(s) for computing the certificates of nonnegativity in Sec.~\ref{sec:algorithms} and their bit complexity analysis in Sec.~\ref{sec:complexity}.
Finally, in Sec.~\ref{sec:sos-perturbations} we present our results 
on approximating a positive polynomial with a sum of squares.

\paragraph{Notations}

Let $\NN$, $\ZZ$, $\QQ$, $\RR$ and $\CC$ be the set of natural numbers
(including zero), integers, rational numbers, real numbers and complex
numbers. We denote by $\K[\X]$ the polynomial ring in the variables
$X_1, X_2, \dots, X_n$ with coefficients in some ring or field $\K$,
where $\K$ can be $\ZZ$, $\QQ$, $\RR$, $\CC$, etc.  We denote by
$\X^{\alpha}$ the monomial $X_1^{\alpha_1}\dots X_n^{\alpha_n}$, where
$\alpha = (\alpha_1, \dots, \alpha_n) \in \NN^n$ is a
multi-index. When $\alpha \in \NN^n$ denotes the exponent of a
monomial, we define $|\alpha| := \sum_i \alpha_i$. For a set of
polynomials $g_1, \dots, g_r \in \K[\X]$, we denote by
$ \langle g_1, \dots, g_r \rangle$ the ideal that they generate. For a
given polynomial $f \in \K[\X]$, we denote by $f^h$ the homogenization
of $f$, i.e.
$f^h \coloneqq \sum_{|\alpha|\leq d} a_{\alpha}
\X^{\alpha}X_0^{d-|\alpha|}\in \K[\X, X_0]$, if
$f = \sum_{|\alpha|\leq d} a_{\alpha} \X^{\alpha} \in \K[\X]$. We also
denote by $f^{(d)}$ the degree $d$ homogeneous part of $f$, it is the
highest degree homogeneous part of $f$ especially when $\deg f = d$.

We use bold lowercase letters to represent a point to distinguish it from the variables, e.g. $\x=(x_1,\dots, x_n) \in \RR^n$. We use $\ntwo{\x }$ to denote the Euclidean norm of $\x$, i.e. $\ntwo{\x } \coloneqq (\sum_{i=1}^{n} x_i^2)^{\frac{1}{2}}$. We denote by $\Sr^n := \{\x \in \RR^{n+1} : \ntwo{ \x } = 1 \}$, the unit sphere in $\RR^{n+1}$.

We denote by $\OO$ and $\OB$ the arithmetic and, respectively, bit
complexity; we also use $\sO$ and $\sOB$, to ignore (poly-)logarithmic factors. For an integer $n$, we denote by $[n]$ the set $\{1, 2, \dots, n\}$. For $r \in \QQ$, we denote by $H(r)$ the height of $r$, which is the maximum absolute value of its numerator and denominator in the representation of $r$ as an irreducible fraction. For $f \in \QQ[\X]$, we denote by $H(f)$ the maximum height of all the coefficients of $f$ and by $\bsz(f)$ the bitsize of $H(f)$, i.e. $\bsz(f) = \lfloor \lg H(f) \rfloor + 1$, where $\lg = \log_2$ and $\lfloor \ \rfloor$  is the floor function that outputs the greatest integer  less than or equal to the input.
We say that $f \in \QQ[\X]$ has size $(d,\tau)$ if it has degree $d$
and bitsize at most $\tau$.

For a polynomial $f \in \RR[\X]$, we denote its nonnegativity over the reals by the abbreviation $f \ge 0$, or $f \ge 0$ on $\RR^n$,
instead of writing  $f(\x) \geq 0$, for all $\x \in \RR^n$.

\section{Preliminaries}
\label{sec:preliminaries}

In this section, we introduce some notations, tools, and results from computational algebraic geometry that we will later use; namely, rational univariate representation, certificates of nonnegativity, lower bounds for critical values of multivariate functions, and coercive polynomials.

\subsection{Ideals and varieties}

Let $\K$ be a field and let $\K[\X]$ be 
the corresponding polynomial ring, where $\X = (X_1, \dots, X_n)$.
Given an ideal $I \subseteq \K[\X]$, we denote by $\sqrt{I} \coloneqq \{f \in \K[\X] : f^n \in I, \text{ for some } n\in \NN\}$ the radical ideal of $I$. For an ideal $I$, if $I=\sqrt{I}$, we say that $I$ is radical. Given an ideal $I$ in $\K[\X]$, we define its variety to be the set $\VV_{\K}(I) \coloneq \{ \x \in \K^n : f(\x) = 0 , \forall f \in \K[\X]\}$. If we omit the subscript we refer to the complex variety, i.e., $\VV(I) = \VV_{\CC}(I)$. 
Given equations $(f_1,\dots,f_r)$, the \emph{solutions ``at infinity''} of the system correspond to the non-zero points in the  variety $\VV_{\C}(f_1^h(0,X_1,\dots,X_n), \dots, f_r^h(0,X_1,\dots,X_n))$, where $f_i^h(0,X_1,\dots,X_n)$ corresponds to the restriction of the homogenization of $f_i$ to the hyperplane $X_0 = 0$.
In particular, we say that the system has \emph{no solutions ``at infinity''}, when {this projective variety is empty}.
For further details on  (computational) commutative algebra and algebraic geometry, we refer the reader to \cite{cox_ideals_2015, cox_david_a_using_2005}.

\subsection{SOS modulo zero-dimensional gradient ideals}
\label{sec:sos-mod-grad-ideal}

Consider a polynomial $f \in \QQ[\X]$. If the infimum of $f$ is attained, then we can decide if $f$ is nonnegative by computing the signs of all its critical values. 
	The latter are the evaluations of $f$ at its critical points,
	that are in turn the (real) zeros of $\Igrad$, the gradient ideal of $f$.
 
\begin{dfn}[Gradient ideal and gradient variety]
\label{def:grad-ideal}
    Given $f \in \R[\X]$, its \emph{gradient ideal}, $\Igrad$, is the ideal generated by 
    all its partial derivatives, that is,
    $
    	\Igrad \coloneq \left\langle \partial_1 f, \dots, 
    	\partial f_n \right\rangle ,
    $    
    where $\partial_i f$ is the derivative of $f$ with respect to $X_i$.
    The \emph{gradient variety} $\VV(\Igrad)$
    consists of all the complex critical points of $f$.
\end{dfn}

Under the assumption that the infimum of $f$ is attained,
Nie, Demmel, and Sturmfels~\cite{nie_minimizing_2006} proved that we can write $f$ as a sum of squares modulo the radical of the gradient ideal of $f$, that is $\sqrt{\Igrad}$. However, their proof is non-constructive.
The most recent result on the constructive version of this
is the \sosrur algorithm by Bender et al.~\cite{btz-sosrur-hal-25},
that computes this SOS decomposition when the gradient ideal is zero-dimensional,
based on a rational univariate representation (RUR) \cite{rouillier_solving_1998} of the points  of the gradient variety.

Given a zero-dimensional variety $V$, we say that a linear form $L \in \QQ[\X]$ is
separating if for every $\bm{p}, \bm{q} \in V$, it holds
$L(\bm{p}) \not= L(\bm{q})$, when $\bm{p} \not= \bm{q}$.

\begin{dfn}[RUR]
    \label{def:rur}
    Consider a zero-dimensional ideal $I \subseteq \Q[\X]$ and a separating linear form $L \in \Q[\X]$ on $\VV(I)$.
    A RUR of $\VVC(I)$ is a tuple of univariate polynomials 
    $R_0(T), R_1(T),\dots,R_n(T) \in \Q[T]$, where $T$ is a new variable, such that
    \[
    \VVC(I)=	\left \{\left(\frac{R_1(t)}{R'_0(t)},\frac{R_2(t)}{R'_0(t)},\cdots, \frac{R_n(t)}{R'_0(t)} \right): t \in \C \textup{ and } R_0(t) = 0\right\},
    \]
    where $R_0$ is square-free and $R'_0$ is its derivative with respect to $T$.
    Also, it holds 
    $t = L\left (\frac{R_1}{R'_0}(t),\dots,\frac{R_n}{R'_0}(t)\right)$, 
    for every root $t \in \C$ of $R_0(T)$.
    We denote this representation as $$\rur(I) = (R_0(T), R_1(T),\dots,R_n(T), T-L(\X)).$$
    We also define an ideal in $\QQ[\X, T]$,
	\[ \Irur \coloneq \ideal{R'_0(T)X_1 - R_1(T), \dots, R'_0(T)X_n - R_n(T)} \subseteq \Q[\X, T] . \]
\end{dfn}

Given $f \in \Q[\X]$ of even degree $d$ and $\rur(\Igrad) =(R_0(T), R_1(T), \cdots, R_n(T), T-\sum_{i=1}^{n} \lambda_i X_i)$ for the gradient ideal of $f$,
\cite[Thm.~3.2]{btz-sosrur-hal-25} (see also \cite[Thm.~4.1]{magron_sum_2023})
proved that, when $f$ attains its infimum and the gradient ideal
$\Igrad$ is zero-dimensional, $f(\X)$ is nonnegative if and only if
the univariate polynomial $(R'_0(t))^d f
\left(\frac{R_1}{R'_0}(t),\dots,\frac{R_n}{R'_0}(t) \right)$ is
nonnegative. This last property can be verified/certified by writing
the univariate polynomial as a weighted sum of squares of polynomials;
see, e.g.,~\cite{bdt-unisos-25}.

\begin{dfn}[The condition $(\Pi)$]
	\label{def:pi}
	Let $f \in \Q[\X]$ be a polynomial of even degree.
	We say that $f$ \emph{satisfies condition $(\Pi)$}, when 
\begin{enumerate}[label=(\arabic*)]
    \item $f$ attains its infimum, and \vspace{-\baselineskip}
    \item[]  \hfill {\Large $(\Pi)$} \vspace{-\baselineskip} 
    \item the gradient ideal $\Igrad$ is zero-dimensional. 
\end{enumerate}
\end{dfn}

\begin{thm}[{\cite[Thm.~3.2]{btz-sosrur-hal-25}; see also \cite[Thm.~4.1]{magron_sum_2023}}]
	\label{thm:SOS-RUR}
    Let $f \in \Q[\X]$ be a polynomial of even degree satisfying $(\Pi)$.
    Let $\rur(\Igrad) =(R_0(T), R_1(T), \cdots, R_n(T), T-\sum_{i=1}^{n} \lambda_i X_i)$ be a rational univariate representation 
  of $\VVC(\Igrad)$. 
   Then, $f$ is nonnegative over $\R^n$ 
   if and only if 
    we can decompose $(R_0')^df$ as an SOS of polynomials modulo the ideal $\Irur \subseteq \Q[\X, T]$,
    that is 
    \begin{equation}
    	\label{eq:sos-rur}
    	(R_0')^df = r + \sum_{i=1}^{n} (R_0' X_i - R_i) q_i = 
    	\sum_{j=1}^{\nu} w_j s_j^2 + \sum_{i=1}^{n} (R_0' X_i - R_i) q_i,
    \end{equation}
    where $r, s_j \in \Q[T]$, $w_j \in \QQ_{+}$, for $j \in [\nu]$,
    and $q_i \in \Q[T, \X]$, for $i \in [n]$.
    \end{thm}

Equation \eqref{eq:sos-rur} leads to a valid certificate of nonnegativity: algebraic identities that, when verified, prove that the polynomial is nonnegative; see \cite[Sec.~4]{btz-sosrur-hal-25}.

\begin{dfn} \label{def:certPos}
    A \emph{certificate of nonnegativity} for $f \in \Q[\x]$ of even degree consists of tuples of  polynomials and constants
    \begin{multicols}{2}
\begin{itemize}
\item $(R_0(T), R_1(T), \cdots, R_n(T)) \in \Q[T]^{n+1}$. \vspace{-.5\baselineskip} 
\item $(q_1,\dots,q_n) \in \Q[T, \X]^n$. \vspace{-.5\baselineskip}
\item $s_1,\dots,s_\nu \in \Q[T]$ (for some $\nu \in \N$).
\end{itemize}
\columnbreak
\begin{itemize}
\item $(\lambda_1,\dots,\lambda_n) \in \Q^n$. \vspace{-.5\baselineskip}
\item $r \in \Q[T]$. \vspace{-.5\baselineskip}
\item $w_1,\dots,w_\nu \in \QQ_{+}$.
\end{itemize}
\end{multicols}
\noindent We say that the certificate is \emph{valid} if,
	\begin{enumerate}[label=(\arabic*)]
        \item $f$  satisfies $(\Pi)$,
        \item $R_0(t)$ is non-zero and square-free, and $T = \sum_{i=1}^n \lambda_i \frac{R_i}{R'_0}(T)$,
        \item We have the inclusion $\left\{ \left(\frac{R_1}{R'_0}(t),\dots,\frac{R_n}{R'_0}(t) \right) : t \in \C, R_0(t) = 0\right\} \supseteq \VV(\Igrad)$, and 
        \item The polynomials satisfy the equality in \eqref{eq:sos-rur}.
    \end{enumerate}
\end{dfn}

\begin{rmk} \label{rmk:dropZeroDim}
    Observe that condition (1) and (3) are related: if condition (3) is satisfied, then the gradient ideal of $f$ must have a finite number of points, so it is zero-dimensional. Therefore, we can replace condition (1) by the condition ``$f$ attains its minimum''.
\end{rmk}

\begin{cor}[{\cite[Cor.~4.4]{btz-sosrur-hal-25}}]
    For any $f \in \Q[\X]$ satisfying $(\Pi)$, we have the equivalence: $f$ is nonnegative if and only if there exists a valid certificate of nonnegativity as in Def.~\ref{def:certPos}.
\end{cor}

If $f$ is not nonnegative, then we can certify this by computing a rational point where $f$ is negative.
\sosrur algorithm computes either of these certificates.

\begin{algorithm}[ht]
    \caption{\sosrur \quad {\cite[Alg.~1]{btz-sosrur-hal-25}}}
    \label{alg:SOS-RUR}

   \textbf{Input:} A multivariate polynomial $f \in \QQ[\X]$ of even degree satisfying $(\Pi)$.

    \vspace{4pt}
  \textbf{Output:} 
  	A \emph{certificate} given by the pair $[\mathtt{NonNeg}, \mathtt{Cert}]$. 
\vspace{-.1\baselineskip}

		\begin{itemize}[leftmargin=1.7cm]
			\item [$\mathtt{NonNeg}$] : a Boolean; \texttt{True} if $f$ is nonnegative,  \texttt{False} otherwise.
			\item [$\mathtt{Cert}$] :  a list containing one of the following (depending on the value of $\mathtt{NonNeg}$):
			\begin{itemize}
        \item If $f$ is not nonnegative ($\mathtt{NonNeg} = \texttt{False}$), then  $\mathtt{Cert} = [\bm{p}]$, \newline
       	where ${\bm p} \in \Q^n$ is such that $f({\bm p}) < 0$.
       	\item If $f$ is nonnegative ($\mathtt{NonNeg} = \texttt{True})$,
  		then  \newline 
  		 $\mathtt{Cert}$ is a \emph{valid certificate} of nonnegativity for $f$; see Def.~\ref{def:certPos} and Eq.~\eqref{eq:sos-rur}.
  		\end{itemize}
	\end{itemize}
\end{algorithm}

\begin{thm}[{\cite[Thm.~3.9,~Cor.~3.18,~Rem.~2.14]{btz-sosrur-hal-25}}]
	\label{thm:sos-rur-complexity}
	Consider $f \in \ZZ[\X]$ of size $(d, \tau)$ and even degree. 
	Assume that $f$ satisfies $(\Pi)$, that is, the gradient ideal of $f$ is zero-dimensional and $f$ attains its infimum.
    The \sosrur algorithm \cite[Alg.~1]{btz-sosrur-hal-25} computes either
    \begin{itemize}
        \item If $f$ is nonnegative, the RUR of $\Igrad$ and the decomposition from Equation \eqref{eq:sos-rur}.
         The largest coefficient of the polynomials involved in this decomposition 
	has bitsize $\sO(d^{2n+1}(d  + n \tau))$ and the total bitsize of the decomposition is $\sO(d^{3n+2}(d  + n \tau))$.
        
        \item If $f$ is not nonnegative, a rational point ${\bm p} \in \Q^n$ such that $f({\bm p}) < 0$. The bitsize of the witness point is $\sO( n d^{2n + 2} (d  + n\tau))$.
    \end{itemize}
    \sosrur is a Monte Carlo algorithm 
    and its bit complexity is
    $\sOB( e^{\omega n + \OO(\lg n)} d^{(\omega +2)n} (n + \tau))$, 
	where $\omega$ is the exponent of matrix multiplication. 	
    When $\Igrad$ has no solutions ``at infinity'', the complexity reduces to $\sOB( e^{\omega n + \OO(\lg n)} \, d^{(\omega +1)n - 1}\, (d + \tau))$.  
    
    The probability of success of \sosrur is $1 - \tfrac{1}{k}$, for any constant $k$.
\end{thm}

\begin{rmk}[Checking zero-dimensionality] \label{rmk:verify0dim}
In this paper we will deal with coercive polynomials, which we will obtain through the stereographic transformation. These polynomials
attain their minimum, but their gradient ideal might not be zero-dimensional.
We can modify the procedures introduced in \cite{btz-sosrur-hal-25} in order to check this last condition within the same complexity of $\sosrur$.
We can perform a deterministic check to verify if a gradient ideal is zero-dimensional and it has no solutions ``at infinity''. To do so, let $(\partial_1f)^h \in \Q[X_0,\dots,X_n]$ be the homogenization of the partial derivative $\partial_i f$.
The resultant of $\langle (\partial_1f)^h,\dots,(\partial_nf)^h, X_0\rangle$ {does not} vanish if an only if the projective variety associated to this ideal is empty, which only happens when the projective variety defined by the homogenized partial derivatives has dimension zero (as otherwise this variety would intersect the hyperplane ``at infinity'' $X_0 = 0$) and they have no solutions ``at infinity''. 
By \cite[Rmk.~2.16]{btz-sosrur-hal-25}, we can perform this check in $\sOB(e^{\omega n} d^{(\omega+1)  n } \tau)$ bit operations, where $\omega$ is the exponent of matrix multiplication.
The    \sosrur algorithm performs several of these resultant computations to compute the RUR of $\Igrad$, so at no extra cost, we can perform this test.
\end{rmk}

The following theorem bounds the bit complexity of verifying if a certificate is valid. For more details on the verification, we refer the reader to \cite[Sec.~4]{btz-sosrur-hal-25}.

\begin{thm}[{\cite[Thm.~4.5]{btz-sosrur-hal-25}}] \label{thm:verifyCertIfPi}
    Consider $f \in \Q[\X]$ of even degree and size $(d,\tau)$. Assume that its infimum is attained (see Rmk.~\ref{rmk:dropZeroDim}). Given a certificate of nonnegativity of $f$ as in Def.~\ref{def:certPos}, we can verify if the certificate is valid in $\sOB( (nd)^{(3\omega+4)n} \, \tau)$ bit operations. 
\end{thm}

\begin{rmk}
    Observe that, a valid certificate of nonnegativity of $f$ exists only if its gradient ideal is zero-dimensional. If this is not the case, then condition (3) in Def.~\ref{def:certPos} is not satisfied.
\end{rmk}

\subsection{Lower bounds for critical values}

In this work we will perturb polynomials without changing their
nonnegativity. In order to quantify how small pertubations we can
consider, we need to study lower bounds for critical values.

\begin{dfn}
	\label{def:glb}
	Let $f \in \RR[\X]$. 
	We denote by $\bd{f}$ the greatest lower bound on the absolute values of the nonzero critical values of $f$, 
	that is 
	\[ \bd{f} := \inf \{|f(\x)| : \x\in \CC^n, f(\x) \neq 0, \nabla f(\x) = \bm{0}\} .
	\] 
\end{dfn}

For polynomial with integer coefficients,
Brownawell and Yap \cite{lower_bounds_Brownawell_Yap_2009}
give a universal lower bound for $\bd{f}$,
that depends on the number of variables, the degree, and the bitsize of the polynomial,
and not on the polynomial itself.

\begin{thm}[{\cite[Cor.~2]{lower_bounds_Brownawell_Yap_2009}}]\label{thm:inf_crit_val_bound}
  Consider any nonzero $f\in \ZZ[\X]$ of degree at most $d$, such that  $H(f) \le 2^\tau$. It holds 
    \[ \bd{f} \geq \bd{n,d,\tau} := ((n+2)^2e^{n+3})^{-(n+1)(n+2)d^{n+1}}(n^n(n+1)\,d \, 2^\tau)^{-(n+1)d^n}
    	= 2^{-\sO(n d^n(n^2 d + \tau))} .
    \]
\end{thm}

\subsection{Coercive polynomials and radius of positivity}
\label{sec:rad-pos}

A (special) family of polynomials that attain their infimum 
and are useful in our study is defined next.

\begin{dfn}[Coerciveness]
	\label{def:coerciveness}
    A function $f : \R^n \to  \R$  is \emph{coercive}, if for any sequence $\{\x_i\}_{i \in \NN}$ of points in $\R^n$, such that $\| \x_i \|_2 \rightarrow +\infty$, we have $f(\x_i) \rightarrow +\infty$. A real polynomial $f \in \RR[\X]$ is coercive if it is coercive as a real function.
\end{dfn}

A coercive function always attains its infimum. For a detailed treatment of the properties of coercive functions, we refer the readers to \cite[Sec. 1.4]{peressini_anthony_l_mathematics_1988}.

\begin{lem}
	\label{lem:coercive-has-min}
    Every coercive function attains its infimum.
\end{lem}

{The radius of positivity of a function (in our case, polynomial) is the radius of a ball centered at the origin}, outside of which the function is nonnegative.
In most cases that we study, we do not need the exact value of the radius, as an upper bound suffices. 
For coercive functions the existence of (trivial or at least effective bounds on) the radius of positivity is guaranteed by the definition of the limit. 

\begin{dfn}[Radius of positivity]\label{def:radius-pos}
	Let $f \in \mathbb{R}[\X]$ be a polynomial in $n$ variables, of even degree. The radius of positivity of $f$ is the smallest nonnegative number $\Rad{f}$ such that $f(\x)\geq 0$, for all $\x \in \RR^n$ with $\ntwo{\x} \geq \Rad{f}$. Abusing the notation, we also use $\Rad{f}$ for an upper bound on this quantity.
\end{dfn}

\section{Stereographic transformation and positivity}\label{sec:steo-transform}

We introduce the \emph{stereographic transformation}.  
This transformation preserves the nonnegativity of polynomials.  
Its principal advantage is that the transformed polynomial is \emph{coercive}, which means that its value tends to \(+\infty\) as the argument tends to infinity.  
Consequently, the transformed polynomial is positive outside a sufficiently large Euclidean ball and all minimizers are confined to its interior.  This property allows us to avoid the difficulty of analyzing and certifying asymptotic critical values of polynomials. 
Moreover, we can derive an \emph{a priori} bound for the radius of such a ball; this bound depends only on the degree of the polynomial and the bitsize of its coefficients, and not on the particular polynomial.
	In this way, we reduce the problem of determining whether a polynomial is globally nonnegative to the problem of verifying its nonnegativity on a compact set that has a size depending solely on its degree and  bitsize.  
The compactness enable us to perform uniformly small perturbations.

\subsection{Stereographic transformation preserves nonnegativity}

To establish nonnegativity of a polynomial, we study transformations that preserve this property.  
We begin with \emph{homogenization}.  
It is a classical result that a polynomial is nonnegative if and only if its homogenization is nonnegative.  We recall the following lemma and refer the reader to Marshall’s monograph~\cite{Marshall2008PositivePA} for the proof.

\begin{lem}[{\cite[Prop.~1.2.4]{Marshall2008PositivePA}}]
\label{lem:aff-hom-equiv}
Let $f\in \RR[\X]$ and let $f^{h}$ 
denote the homogenization of $f$.  
Then, $f$ is nonnegative if and only if $f^{h}$ is nonnegative.
\end{lem}

From a geometric point of view, the \emph{stereographic projection} maps the sphere, minus the projection point, diffeomorphically onto a hyperplane. Within its domain it is smooth, bijective, and conformal, although it does not preserve distances or areas.

We need a variant of stereographic projection, that we call the \emph{stereographic transformation}. 
We obtain it from substituting into the homogenized polynomial the numerator(s) of the stereographic projection;
this results in a polynomial and helps us to avoid manipulating rational functions.

\begin{dfn}[Stereographic transformation]\label{dfn:steo-trans}
Consider $f \in \RR[\X]$ with degree $d$ and let $f^{h}$ denote its homogenization.  
The \emph{stereographic transformation} of $f$ is 
\[
	\nsteo{f}(\mathbf{X}) \coloneq 
f^{h}\bigl(2X_{1},\dots,2X_{n},\,-1+\sum\nolimits_{i=1}^{n}X_{i}^{2}\bigr).
\]
\end{dfn}

As a function, the stereographic transformation of a polynomial behaves as follows,
        \begin{align} \label{eq:steoCases}
    \nsteo{f}(\x)
   =  \begin{cases}
        (-1+\ntwo{\x}^2)^d \ f\bigl(\frac{2x_1}{-1+\ntwo{\x}^2}, \dots, \frac{2x_n}{-1+\ntwo{\x}^2}\bigr),  & 
        \text{ if } \ntwo{\x} \neq 1 
        \\
        2^d \ f^{(d)}(\x), & 
        \text{ if } \ntwo{\x} = 1
        \end{cases}, 
    \end{align}
where $f^{(d)}$ is the highest degree homogeneous part of $f$.

We now show that the stereographic transformation preserves nonnegativity of polynomials.

\begin{thm}\label{thm:aff-steo-equiv}
$f \in \RR[\X]$ is nonnegative on $\RR^n$ if and only if $\nsteo{f} \in \RR[\X]$ is nonnegative on $\RR^n$.
\end{thm}
\begin{proof}
If $f\in \RR[\X]$ is homogeneous in variables $X_1,\dots, X_n$, then
\[
\nsteo{f}(\mathbf{X})
= f^{h}\bigl(2X_{1},\dots,2X_{n},-1+\textstyle\sum_{i=1}^{n}X_{i}^{2}\bigr)
= f\bigl(2X_{1},\dots,2X_{n}\bigr),
\]
and the claim follows immediately.  
Hence, we may assume that $f$ is not homogeneous.

The proof proceeds by successively reducing the nonnegativity of $f$ to that of $f^{h}$ on the unit sphere, and finally to $\RR^{n}$ via the stereographic projection, which establishes the equivalence for $\nsteo{f}$. We have
\begin{align*}
f \ge 0 \text{ on } \RR^{n}
\iff\;& f^{h} \ge 0 \text{ on } \RR^{n+1} 
\iff f^{h} \ge 0 \text{ on the unit sphere } \mathbb{S}^{n}
\\[0.5ex]
\iff\;& f^{h} \ge 0 \text{ on } \mathbb{S}^{n}\setminus\{p\},
\quad\text{where } p=(0,\dots,0,1) \text{ is the north pole of } \mathbb{S}^{n},
\\
&\quad\text{(by the continuity of } f^{h} \text{)}
\\[0.5ex]
\iff\;&
f^{h}\!\left(
  \frac{2X_{1}}{1+\sum_{i=1}^{n}X_{i}^{2}},\,
  \dots,\,
  \frac{2X_{n}}{1+\sum_{i=1}^{n}X_{i}^{2}},\,
  \frac{-1+\sum_{i=1}^{n}X_{i}^{2}}{1+\sum_{i=1}^{n}X_{i}^{2}}
\right) \ge 0 \text{ on } \RR^{n}
\\[0.5ex]
\iff\;&
f^{h}\bigl(2X_{1},\dots,2X_{n},-1+\textstyle\sum_{i=1}^{n}X_{i}^{2}\bigr) \ge 0 \text{ on } \RR^{n},
\\
&\quad\text{(since } 1+\sum_{i=1}^{n}X_{i}^{2} > 0 \text{ everywhere, we can omit the denominators)}
\\[0.5ex]
\iff\;& \nsteo{f} \ge 0 \text{ on } \RR^{n}.
\qedhere
\end{align*}
\end{proof}

\subsection{Stereographic transformation and coercive polynomials}

In what follows we show that, when the constant term of $f$ is positive, $\nsteo{f}$ is a coercive function (Def.~\ref{def:coerciveness}).
This follows from the observation that the highest degree homogeneous part of $\nsteo{f}$ is of even degree on each variable and each leading monomial has a positive coefficient. 

\begin{thm}\label{thm:coercive-direct-steo}
   Let $f \in \ZZ[\X]$ be a polynomial of even degree $d$ with $f(\bm{0}) > 0$.  When $\ntwo{\x} \geq 1 $,  we have
 \[
 \nsteo{f}(\x) \geq  \ntwo{ \x }^{2d} -  2^{n+2d}\ H(f)\ \ntwo{ \x }^{2d-1} .
 \]
   Thus, $\nsteo{f}$ is coercive and attains its infimum.
\end{thm}

\begin{proof}
    Let $f = \sum_{|\alpha| \leq d} a_{\alpha}\X^{\alpha}$ and denote by $f^h = \sum_{|\alpha| \leq d} a_{\alpha}\X^{\alpha}X_0^{d-|\alpha|}$ its homogenization. For $\x \in \RR^n$, we have
    \begin{align}
        \nsteo{f}(\x) & =   a_{\bm{0}} \ntwo{ \x }^{2d} + a_{\bm{0}} \left ( \sum_{i=1}^{d}\binom{d}{i}(-1)^i \ntwo{ \x }^{2d-2i} \right )+ \sum_{0<|\alpha| \leq d} a_{\alpha}2^{|\alpha|}\x^{\alpha}(-1+\ntwo{ \x }^2)^{d-|\alpha|} 
\label{eq:steo-expansion} \\ \nonumber
        & \geq a_{\bm{0}} \ntwo{ \x }^{2d} - \left| a_{\bm{0}} \left ( \sum_{i=1}^{d}\binom{d}{i}(-1)^i \ntwo{ \x }^{2d-2i} \right )\right| -  \left| \sum_{0<|\alpha| \leq d} a_{\alpha}2^{|\alpha|}\x^{\alpha}(-1+\ntwo{ \x }^2)^{d-|\alpha|}\right| .
    \end{align}
    Observe that, for any $\alpha$, $|a_\alpha| \leq H(f)$, as $a_\alpha \in \Z$.
    Moreover, $|\x^{\alpha}| \leq \ntwo{\x}^{|\alpha|}$. As $\ntwo{ \x } \geq 1$, $\ntwo{ \x }^i < \ntwo{ \x }^j$, for $i < j$, and  $ \ntwo{ \x }^{|\alpha|}(-1+\ntwo{ \x }^2  )^{d-|\alpha|} 
        \leq \ntwo{\x}^{2d - |\alpha|}$. 
        Hence, we have that    
\begin{align*}      \left| a_{\bm{0}} \left( \sum_{i=1}^{d}\binom{d}{i}(-1)^i \ntwo{ \x }^{2d-2i} \right) \right| &   \leq 2^{d} \ H(f)  \ \ntwo{ \x }^{2d-2} \qquad\qquad  \text{ and }\\ 
 \left| \sum_{0<|\alpha| \leq d} a_{\alpha}2^{|\alpha|} \x^{\alpha}(-1+\ntwo{ \x }^2)^{d-|\alpha|} \right| &   \leq
2^d  \left( {d + n \choose n} - 1 \right)\ H(f) \ \ntwo{\x}^{2d - 1} .
\end{align*}
Observe that, as $a_{\bm{0}} \in \Z$ is positive, $a_{\bm{0}} \geq 1$. Therefore,
   \begin{align} \label{eq:ineqSteo}
        \nsteo{f}(\x) \geq \ntwo{ \x }^{2d} - 2^d \ {d + n \choose n} \ H(f) \ \ntwo{\x}^{2d - 1} \geq
        \ntwo{ \x }^{2d} - 2^{n + 2d} \ H(f) \ \ntwo{\x}^{2d - 1} .
    \end{align}
        In particular, when $\ntwo{\x} \geq 2^{n+2d}H(f)$, we have $\nsteo{f} \geq 0$. It also follows that $\nsteo{f}$ is coercive and therefore attains infimum.
\end{proof}

\begin{cor}\label{cor:steo-radius-estimates}
    Let $f \in \ZZ[\X]$ be a polynomial of even degree $d$ such that $f(\bm{0}) > 0$ and $H(f) \leq  2^\tau$.
    Let 
    \begin{align}\label{eq:defOfRad}
        \Rad(n,d,\tau) := 2^{n+2d+\tau} = 2^{\OO(n + d + \tau)} \enspace.        
    \end{align}
     If $\ntwo{\x} \geq \Rad(n,d,\tau)$, then $\nsteo{f}(\x) \geq 0$.
    We also use $\Rad(n,d,\tau)$ for the value of $\Rad(\nsteo{f})$ following Def.~\ref{def:radius-pos}.
\end{cor}

\begin{rmk}
   We can prove the coerciveness of $\nsteo{f}$ and compute (actually bound) the radius $\Rad$ in a more general framework, that is  the class of \emph{stably bounded from below} polynomials, studied by Marshall~\cite[Thm.~5.1]{marshall2003optimization}, see also~\cite[Lem.~1]{jibetean2005semidefinite}. 
    We choose to prove the coerciveness directly for the polynomial $\nsteo{f}$, as this results in a better bound of the radius $\Rad(\nsteo{f})$. 
 In particular, following Marshall~\cite[Thm.~5.1]{marshall2003optimization} or Jibetean and Laurent~\cite[Lem.~1]{jibetean2005semidefinite} 
 we bound the radius by $\max \{ 1, \frac{1}{\min \nsteo{f}^{(2d)}} \sum_{|\alpha| \leq 2d-1} |c_{\alpha}|\}$, where $c_{\alpha}$ is the coefficient of $\X^{\alpha}$ in the polynomial $\nsteo{f}$, $\nsteo{f}^{(2d)}$ is the highest degree homogeneous part of $\nsteo{f}$, and $\min \nsteo{f}^{(2d)}$ is its minmum on the unit sphere $\Sr^{n-1}$.  
 Straightforward computations yield the estimations
 $\min {\nsteo{f}}^{(2d)} =a_0 \geq 1$ and 
 $\sum_{|\alpha| \leq 2d-1} |c_{\alpha}| 
 \leq (n+1)^{d}\,2^{\,2n+4d +\tau},$
 and so the radius is bounded by $(n+1)^d 2^{2n+4d+\tau}$.
\end{rmk}

\subsection{Tracing witness points from {$\nsteo{f}$} to $f$}

We now establish a relation between a negative evaluation of $\nsteo{f}$ and a corresponding negative evaluation of $f$. 
In particular, any (rational) point at which $\nsteo{f}$ takes a negative value corresponds to an explicit (rational) witness point; the latter establishes that $f$ is not nonnegative.

\begin{lem}\label{lem:stereo-witness}
Let $f = \sum_{\bm{\alpha}} a_{\alpha} X^{\bm{\alpha}} \in \ZZ[\X]$ be a polynomial of even degree $d$, and let $\x \in \QQ^{n}$ 
be such that $\nsteo{f}(\x) < 0$. 
Define $H(\x) := \max_{1\le i\le n} H(x_{i})$ and 
\[
C_{f,\x} := \,2^{\,n+2d-1} \ H(\x)^{(n+1)d-1} \ H(f).
\]
We define
\[
\mathscr{Q}(f,\x)
 := \begin{cases}
    \left (\frac{2x_1}{-1+\ntwo{\x}^2 },\dots,\frac{2x_n}{-1+\ntwo{\x}^2} \right), \qquad & \text{ if } \ntwo{\x} \neq 1  \\
    ({2x_1}C_{f, \x},\dots,{2x_n}C_{f, \x}), \qquad & \text{ if } \ntwo{\x} = 1 . 
\end{cases}
\]
Then $f(\mathscr{Q}(f,\x)) < 0$.
Moreover, we have
\[
H(\mathscr{Q}(f,\x)) \leq 2^{n+2d} \, H(\x)^{(n+1)d}H(f).
\]

\end{lem}

\begin{proof}
Given the expansion of the stereographic transformation from Eq.~\ref{eq:steo-expansion}, we distinguish two cases.

\noindent
\emph{Case 1: $\ntwo{\x} \ne 1$.}
From \eqref{eq:steo-expansion} we can factor out $(-1+\ntwo{\x}^{2})^{d}$, which leads to
\[
\nsteo{f}(\x)
= (-1+\ntwo{\x}^{2})^{d}
  \Bigl(f_{\bm 0}+\sum_{0<|\alpha|\le d}
   f_{\alpha}\,2^{|\alpha|}\,\x^{\alpha}\,(-1+\ntwo{\x}^{2})^{-|\alpha|}\Bigr).
\]
Since $d$ is even, the factor $(-1+\ntwo{\x}^{2})^{d}$ is nonnegative.
Hence, the sign of $\nsteo{f}(\x)$ is determined by the factor in the parentheses.  
The assumption $\nsteo{f}(\x)<0$ implies
\[
f_{\bm 0}+\sum_{0<|\alpha|\le d}
   f_{\alpha}\,2^{|\alpha|}\,\x^{\alpha}\,(-1+\ntwo{\x}^{2})^{-|\alpha|} <0,
\]
which is precisely the evaluation $f(\mathscr{Q}(f,\x))<0$ with
$\mathscr{Q}(f,\x)=\bigl(2x_{1}/(-1+\ntwo{\x}^{2}),\dots,2x_{n}/(-1+\ntwo{\x}^{2})\bigr)$. In this case, we have $H(\mathscr{Q}(f,\x)) \leq (n+1)H(\x)^{2n}$.

\medskip

\noindent
\emph{Case 2: $\ntwo{\x}=1$.}
In this case, \eqref{eq:steo-expansion} simplifies to
\begin{equation}
\label{eq:steo-x1}
	\nsteo{f}(\x) = \sum_{|\alpha|=d} f_{\alpha}\,2^{|\alpha|}\,\x^{\alpha}= 2^d\sum_{|\alpha|=d} f_{\alpha}\,\x^{\alpha},
\end{equation}
since all terms with $|\alpha|<d$ vanish.
As $\nsteo{f}(\x)<0$ and $f_{\alpha}\in\ZZ$, we have the lower bound
$\nsteo{f}(\x) \le -\,{H(\x)^{-nd}}.$ 
Because $d$ is even, the sign of 
$f(\mathscr{Q}(f,\x))
= (C_{f,\x})^{d}\,
  f^{h}\bigl(2x_{1},\dots,2x_{n},\,1/C_{f,\x}\bigr)$
coincides with the sign of
$f^{h}\bigl(2x_{1},\dots,2x_{n},\,1/C_{f,\x}\bigr)$.  
We thus estimate
\begin{align*}
f^{h}\bigl(2x_{1},\dots,2x_{n},\,1/C_{f,\x}\bigr)
&= \sum_{|\alpha|\le d} a_{\alpha}\,2^{|\alpha|}\,\x^{\alpha}\,C_{f,\x}^{-(d-|\alpha|)}= \nsteo{f}(\x)
   + \sum_{|\alpha|<d} a_{\alpha}\,2^{|\alpha|}\,\x^{\alpha}\,C_{f,\x}^{-(d-|\alpha|)} \qquad\text{(by \eqref{eq:steo-x1})}\\
&\le -\frac{1}{H(\x)^{nd}}
   + \sum_{|\alpha|<d} H(f)\,2^{\,d-1}\,H(\x)^{\,d-1}\,C_{f,\x}^{-1} \qquad\qquad \quad\text{(since }|\x^{\alpha}|\le H(\x)^{|\alpha|})\\
&\le -\frac{1}{H(\x)^{nd}}
   + 2^{\,n+d-1} H(f)\,2^{\,d-1} H(\x)^{\,d-1}\,C_{f,\x}^{-1} 
 %
\quad < \quad 0.
\end{align*}
Therefore $f(\mathscr{Q}(f,\x)) = (C_{f,\x})^{d}\, f^{h}\bigl(2x_{1},\dots,2x_{n},\,1/C_{f,\x}\bigr)<0$, and  $H(\mathscr{Q}(f,\x)) \leq H(\x)^{(n+1)d}H(f)2^{n+2d}$.
\end{proof}

\subsection{Bounds on critical values}
In this subsection, we study upper bounds for the smallest absolute value of the nonzero critical values of $\nsteo{f}$.
For this, we need the following auxiliary inequality relating the coefficient sizes of $f$ and $\nsteo{f}$. 

\begin{lem}
\label{thm:steo-coef-size}
    Let $f \in \ZZ[\X]$ 
    be a polynomial in $n$ variables of degree $d$ such that $H(f) \leq 2^{\tau}$. Then, 
    \[
    	H(\nsteo{f}) \leq 
    	 2^d  \, \binom{n+d}{d} \, (n+1)^d \, H(f)  
         \leq 
    	2^{d\lg(n+1)+ n+2d+\tau}.
    \]
\end{lem}

\begin{proof}
    If we write $f = \sum_{|\alpha| \leq d} f_{\alpha} \X^{\alpha}$, we have that
    $
    \nsteo{f}(\X) 
    = \sum_{|\alpha| \leq d} a_{\alpha}2^{|\alpha|}\X^{\alpha}(-1+\sum_{i=1}^{n}X_i^2)^{d-|\alpha|}.
    $
    To estimate $H(\nsteo{f})$ we bound the coefficients of the right-hand side
    of the previous equation. Thus
    \begin{align*}
    H(\nsteo{f})  & \leq 2^d \, H(f)  \sum_{|\alpha| \leq d} H \left( \X^{\alpha}(-1+\sum_{i=1}^{n}X_i^2)^{d-|\alpha|} \right)   \\  
     & \leq 2^d \, H(f) \, \sum_{|\alpha| \leq d} \max \left \{\binom{d}{k_0,\dots,k_n} : k_0+\dots+k_n = d \right \}    \\
     & \leq 2^d  \, \binom{n+d}{d} \, (n+1)^d \, H(f) \enspace \qquad
       \text{\Big(since $\binom{d}{k_0,\dots,k_n} \leq (n+1)^d$\Big)}
     \qedhere
    \end{align*}
    
\end{proof}

\begin{cor} \label{cor:boundCritValSteo}
    Let $f \in \ZZ[\X]$ 
    be a polynomial in $n$ variables of degree $d$ such that $H(f) \leq 2^{\tau}$. 
    The smallest absolute value of the nonzero critical values of $\nsteo{f}$, $\bd(\nsteo{f})$, is upper-bounded by 
    \[
    	\bd{\nsteo{f}} \geq \bd{n,  2d, d\lg(n+1)+ n+2d+\tau}
    	= 2^{-\sO(n (2d)^n(n^2 d + \tau))} ,\]
    where $\bd{n,  d,\tau}$ is as defined in Thm.~\ref{thm:inf_crit_val_bound}.
\end{cor}

\begin{proof}
        By Lem.~\ref{thm:steo-coef-size}, we have $H(\nsteo{f}) \leq 2^{d\lg(n+1)+ n+2d+\tau }$. Combining this bound with Thm.~\ref{thm:inf_crit_val_bound}, we have $\bd{\nsteo{f}} \geq \bd{n,  2d, d\lg(n+1)+ n+2d+\tau}$.
\end{proof}

\section{Perturbations preserving nonnegativity}\label{sec:HJ-pert}

The stereographic transformation allows us to transform (back-and-forth) nonnegative polynomials into nonnegative polynomials attaining their infimum.
For a polynomial attaining its infimum, we can certify its nonnegativity
by certifying that all its critical values are nonnegative.
Algorithm \sosrur can do this when the gradient ideal is
zero-dimensional (and so, there is a finite number of critical
points).
We introduce a perturbation that guarantees that the
gradient ideal of the perturbed polynomial is zero-dimensional.

\subsection{Hanzon-Jibetean perturbations}

Hanzon and Jibetean~\cite{hanzon2003global} studied algebraic algorithms to approximate the global minimum of a polynomial. Their main idea is, when the critical points are finite, to approximate the critical values by approximating these points. However, when this is not the case, then they showed that it is possible to perturb the input polynomial
to obtain a coercive polynomial that has a similar minimum but also a finite number of critical points. Hence, to approximate the minimum, they proceed by considering successive smaller perturbations.

Inspired by their construction, we show that, given a coercive polynomial, one can construct a sufficiently small perturbation, which we call \emph{Hanzon-Jibetean perturbation}, such that the perturbed polynomial is nonnegative if and only if the original polynomial is nonnegative.

\begin{dfn}[HJ perturbation]
	\label{def:pert-lambda}
	Let $f \in \RR[\X]$ be a polynomial in $n$ variables, of degree $d$,
	and let $\lambda$ be any positive real number.
	The positive (respectively, negative) HJ perturbation of $f$ is defined as
	\[
		\pospert{f}(\X) \coloneqq f(\X) 
		+ \lambda \sum\nolimits_{i=1}^{n} \left(1+X_i^2 + X_{i}^{d+2} \right)\,, \text{ and} 
	\] 
	\[
		\negpert{f}(\X) \coloneqq f(\X) 
		- \lambda \sum\nolimits_{i=1}^{n}(1+X_i^2 + X_{i}^{d}) .
	\] 
\end{dfn}
Notice that $\pospert{f}$ and $\negpert{f}$ apply perturbations of different degrees.
The slightly higher degree perturbation in $\pospert{f}$ simplifies 
some arguments in the proof that follows and it corresponds to
(a slight variant of) 
the original perturbation introduced by Hanzon-Jibetean.

\begin{thm}[{\cite[Sec.~3 and Prop.~3.1]{hanzon2003global}}]
 \label{thm:0-dim-plus-pert}
     Let $f \in \R[\X]$ be a polynomial of even degree $d$. 
     For any positive constant $\lambda \in \R$, the polynomial 
     $\pospert{f}(\X)$ is coercive and its corresponding gradient ideal $I_{\nabla \pospert{f}}$ is zero-dimensional.
     Moreover, the gradient ideal has $(d+1)^n$ complex solutions (counted with multiplicities) and no solutions ``at infinity''.
 \end{thm}

\begin{proof}
    Following \cite[Prop. 3.1]{hanzon2003global}, the partial derivatives of $\pospert{f}$ form a \GB  basis for the gradient ideal
    	$I_{\nabla \pospert{f}}$ with respect to any graded order. Each partial derivative has leading monomial $X_i^{d+1}$, so the stair-case of the \GB basis is closed, and thus, the gradient ideal is zero-dimensional. Moreover, $\K[\X] / I_{\nabla \pospert{f}}$ has dimension $(d+1)^n$, so the 
    	gradient system has the same number of complex points, counted with multiplicities. Each partial derivative at infinity is a power of a monomial, that is, $(\partial_i f)^h(0,X_1,\dots,X_n) =  X_i^{d+1}$, so there are no solutions ``at infinity''.
\end{proof}

We show that, for generic choices of $\lambda$, 
that is for all $\lambda$ belonging to a Zariski open set, 
the perturbed polynomials of Def.~\ref{def:pert-lambda} have radical
gradient ideal, a zero dimensional gradient variety, and no solutions
``at infinity''.

\begin{prop}
\label{prop:0-dim-rad-pert-generically}
    For a generic  $\lambda \neq 0$, $\pospert{f}$ and $\negpert{f}$ have zero-dimensional and radical gradient ideals.
    Moreover, the gradient system of $\pospert{f}$, respectively $\negpert{f}$, has $(d+1)^n$, respectively $(d-1)^n$, complex solutions (counted with multiplicities) and no solutions at ``infinity''. 
\end{prop}

\begin{proof}
   We show the statement for $F_t =\sum_i (1+X_i^2 + X_{i}^{\,l}) + t f$, where $l=d$ or $d+2$. The original claim then follows by taking $t=\pm 1/\lambda$, $\lambda\neq 0$. 
    We now establish zero-dimensionality of $I_{\nabla F_t}$ for generic $t\neq 0$ and treat only the case $l=d$, as $l =d +2$ was treated in Thm.~\ref{thm:0-dim-plus-pert}. 

For $t=0$ we have a polynomial $\sum_i (1+X_i^2 + X_{i}^{\,d})$, whose gradient ideal is obviously zero-dimensional. 
    Consider the variety 
    \begin{align*}
        C\ =\ \left\{({\bf a}, t)\in \mathbb{A}^n_{\mathbb{C}}\times \mathbb{A}^1_{\mathbb{C}}: 
        \partial_1 F_t({\bf a})=\dots=\partial_n F_t({\bf a})=0\right\}
    \end{align*}
and the projection $\pi|_C: C\rightarrow \mathbb{A}_{\mathbb{C}}^1$ to the second argument. By \cite[Thm. 3.1.5]{EGAIV3} the dimension of $C\cap \pi^{-1}(t)$ is an upper-semicontinuous function of $t\in \mathbb{A}^1_{\mathbb{C}}$. Therefore, since $C\cap \pi^{-1}(0)$ is a finite set, so is a generic fiber $C\cap \pi^{-1}(t)$, $t \in \mathbb{A}^1_{\mathbb{C}}\setminus \{0\}$. The latter exactly means that the gradient ideal of $F_t$ is zero-dimensional.
As at $t = 0$ there are not solutions ``at infinity'', the same holds for a generic fiber, and so the number of (affine) solutions for a generic fiber is the B\'ezout bound $(d-1)^n$.
    
Radicality of a zero-dimensional ideal $I_{\nabla F_t}$ (whose zero set is denoted by $X_t\subset \mathbb{A}^n_{\mathbb{C}}$) is equivalent to reducedness of local rings $\mathcal{O}_{X_t,{\bf a}}$ at every ${\bf a} \in X_t$. The latter condition is equivalent for ${\bf a} \in X_t$ to be a non-denegerate critical point of $F_t$, that is, 
$\partial_1 F_{t}({\bf a})=\dots=\partial_n F_{t}({\bf a})=0$ and 
$\det\left(\frac{\partial^2 F_{t}}{\partial x_i \partial x_j}({\bf a})\right)\neq 0$. 
Consider now the variety 
\begin{align}
D\ =\ \left\{({\bf a},t)\in \mathbb{A}_{\CC}^n\times \mathbb{A}^1_{\CC}: \partial_1 F_{t}({\bf a})=\dots=\partial_n F_{t}({\bf a})=0,\ \det\left(\frac{\partial^2 F_{t}}{\partial x_i \partial x_j}({\bf a})\right) = 0 \right\}
\end{align}
of ``bad'' pairs (that is, non-reduced points together with the parameter $t$).
Its projection $\pi(D)$ under $\pi: \mathbb{A}_{\CC}^n\times \mathbb{A}^1_{\CC}\rightarrow \mathbb{A}^1_{\CC}$ is a constructible subset of $\mathbb{A}^1_{\CC}$.
Because $0\notin\pi(D)$ (as the gradient ideal of $\sum_i (1+X_i^2 + X_{i}^{l})$ is radical), $\pi(D)$ is a finite set.
This means that $I_{\nabla F_t}$ is radical for all but finitely many $t\in \mathbb{A}^1_{\CC}$.  
\end{proof}

Working with polynomials with zero-dimensional gradient ideal and no solutions ``at infinity'' allow us to compute our certificates faster, see Thm.~\ref{thm:sos-rur-complexity}. Prop. \ref{prop:0-dim-rad-pert-generically} implies that for most $\lambda$'s, $\negpert{\nsteo{f}}$ satisfies the above assumptions. In what follows we quantify how many \emph{unlucky} choices of $\lambda$'s there are.
    
\begin{lem} \label{lem:unluckyChoicesNegPert}
    Consider non-zero $f \in \ZZ[\X]$, in $n$ variables, of even degree $d$, such that $f$ \emph{and} its highest degree homogeneous part are positive. 
    We have that there are at most $(2d - 1)^{n-1} (2d - 1 + n)$ values for $\lambda$ such that either the gradient ideal of  $\negpert{\nsteo{f}}$ is not zero-dimensional or it has solutions ``at infinity''.
\end{lem}
\begin{proof}
    We proceed as in Rmk.~\ref{rmk:verify0dim}. Let $g_i := \partial_i(\negpert{\nsteo{f}})$ and consider its homogenization $g_i^h \in \ZZ[X_0,X_1,\dots,X_n]$.
    The resultant of the homogeneous system $(g_1^h, \dots,
    g_n^h,X_0)$ vanishes if and only if this system has a solution in
    $\PP^n$, the $n$-dimensional complex projective space.
    If the resultant does not vanish, then the system has no solutions at infinity. Moreover, the variety defined by $(g_1^h, \dots, g_n^h)$ is zero-dimensional, as otherwise it has to intersect the hyperplane $X_0 = 0$.
The resultant of this system is a polynomial in $\lambda$ of degree at most $(2d - 1)^{n-1} (2d - 1 + n)$ \cite{cox_david_a_using_2005}. The proof follows straightforwardly by noting that, by Prop.~\ref{prop:0-dim-rad-pert-generically}, this resultant is a non-zero polynomial. 
\end{proof}

\subsection{Positive perturbations preserving nonnegativity}

We quantify the size of perturbations of coercive polynomials
that preserve nonnegativity.

\begin{thm}
	\label{thm:f-pos-equiv-pertf-pos}
    Let $f \in \ZZ[\X]$ be a coercive polynomial in $n$ variables, of even degree $d$. Consider   $\lambda \in \big(0,\frac{1}{6n}\frac{\bd{f}}{\Rad{f}^{d+2}}\big]$, where $\bd(f)$ is a lower bound on the absolute values of the nonzero critical 
	values of $f$ (Def.~\ref{def:glb})
	and $\Rad(f)$ is the radius of positivity
	(Def.~\ref{def:radius-pos}). We assume with no loss of generality that $\Rad(f) > 1$.
    We have the following equivalence:
    \[
        f \ge 0 \iff \pospert{f} \ge 0 .
    \]
\end{thm}

\begin{proof}
The implication ($\implies$) is immediate.
The perturbation term
$
\lambda \, (
\sum_{i=1}^{n} 1 + X_i^2 + X_{i}^{\,d+2})
$
is nonnegative on $\RR^n$ because $d$ is even and $\lambda > 0$.
Hence, if $f(\x)\ge 0$ for $\x\in\RR^{n}$,
then $\pospert{f}(\x)\ge 0$.

For the converse implication ($\impliedby$),
suppose that $f$ is not nonnegative.
Let
$
\disc( \Rad{f}) \coloneqq \{\x\in\RR^{n}:\|\x\|\le \Rad{f}\}
$
and choose $\x_{0}=(x_{01},\dots,x_{0n})\in \disc(\Rad{f})$,
such that $f(\x_{0})=\min_{\x\in \disc(\Rad{f})} f(\x).$
Since $f$ is not nonnegative, $f(\x_{0})<0$. By the definition of $\disc(\Rad{f})$, 
we have $\x_0 \in \textup{Int}\,\disc(\Rad{f})$ and $\x_0$ is a critical point of $f$, 
where $\textup{Int}\,\disc(\Rad{f})$ is the interior of the ball $\disc(\Rad{f})$.
As $\bd{f}$ is a positive lower bound for the absolute values
of all nonzero critical values of $f$, it follows that
$f(\x_{0})\le -\,\bd{f}.$
Moreover,
\[
\lambda \sum_{i=1}^{n} (1+x_{0i}^2+x_{0i}^{\,d+2}) \leq \frac{1}{6n}\frac{\bd{f}}{\Rad{f}^{d+2}}
\sum_{i=1}^{n} (1+x_{0i}^2+x_{0i}^{\,d+2})
\le
\frac{1}{6n}\frac{\bd{f}}{\Rad{f}^{d+2}}
\cdot 3n \, \Rad{f}^{d+2}
= \frac{1}{2}\,\bd{f}.
\]
The evaluation of $\pospert{f}$ at $\x_{0}$ yields
$
\pospert{f}(\x_{0})
\le -\,\bd{f}+\tfrac{1}{2}\bd{f}
= -\tfrac{1}{2}\bd{f}
<0,
$
which contradicts the assumption that $\pospert{f}$ is nonnegative on $\RR^n$.
Thus, $f$ is nonnegative on $\RR^{n}$.
\end{proof}

As the stereographic transformation preserves nonnegativity
(Thm.~\ref{thm:aff-steo-equiv}) and leads to coercive polynomials, we
can mix the previous theorem with this transformation. In what follows
we do so by mixing these ideas with precise bounds for critical values
and radii of positivity.

\begin{thm}
	\label{thm:f-pos-effect-equiv-pert-steo}
    Let $f\in \RR[\X]$ be a polynomial in $n$ variables, of even degree $d$, such that $f(\bm{0}) > 0$ and $H(f) \leq 2^{\tau}$.
    Consider the following positive $\epsilon(n,d,\tau)$ defined as
    \begin{align} \label{eq:constantPert}
        \epsilon(n,d,\tau) \coloneqq \frac{1}{6n}\frac{\bd{n,  2d, d\lg(n+1)+ n+2d+\tau}}{\Rad{n,d,\tau}^{2d+2}}
        =  2^{-\sO(n (2d)^n(n^2 d + \tau))} ,
    \end{align}
    where $\bd{n,  d,\tau}$, respectively $\Rad{n,d,\tau}$, are as defined in Thm.~\ref{thm:inf_crit_val_bound}, respectively Cor.~\ref{cor:steo-radius-estimates}.
         Then, for any $\lambda \in (0, \epsilon(n,d,\tau)]$, we have, over $\R^n$,
    \[
    f \geq 0 \iff
     \pospert{\nsteo{f}} \geq 0,
    \]
    Moreover $\pospert{\nsteo{f}}$ satifies $(\Pi)$ and, for generic choices of $\lambda$, the gradient ideal of $\pospert{\nsteo{f}}$ is also radical.
\end{thm}

\begin{proof}
First, we will prove that the transformation preserves nonnegativity.
By Thm.~\ref{thm:aff-steo-equiv}, $f \geq 0 \iff \nsteo{f} \geq 0$. Hence, we will prove that $\nsteo{f} \geq 0 \iff \pospert{\nsteo{f}} \geq 0$.
By Cor.~\ref{cor:boundCritValSteo}, we have $\bd{\nsteo{f}} \geq \bd{n,  2d, d\lg(n+1)+ n+2d+\tau}$. By Cor.~\ref{cor:steo-radius-estimates}, we can choose $\Rad{\nsteo{f}} =  \Rad{n,d,\tau}$. 
As $\nsteo{f}$ is a coercive (Thm.~\ref{thm:coercive-direct-steo}),
the equivalence $\nsteo{f} \geq 0 \iff \pospert{\nsteo{f}} \geq 0$
follows straightforwardly from Thm.~\ref{thm:f-pos-equiv-pertf-pos}.

To conclude, observe that $\nsteo{f}$ is coercive (and hence also $\pospert{\nsteo{f}}$ for all $\lambda>0$) and, by Thm.~\ref{thm:0-dim-plus-pert}, the gradient ideal of $\pospert{\nsteo{f}}$ is zero-dimensional, so it satisfies condition $(\Pi)$. Moreover, by Prop.~\ref{prop:0-dim-rad-pert-generically}, for a generic $\lambda$, the gradient ideal of $\pospert{\nsteo{f}}$ is also radical.
\end{proof}

\subsection{Negative perturbations preserving nonnegativity}
Given any polynomial $f$, we can always can find $\lambda > 0$ such that $\pospert{f}$ is nonnegative, regardless of $f$. This is why, if we want positive perturbations $\pospert{}$ to preserve positivity, we are forced to work with very small values of $\lambda$, depending on the size of $f$ and not the function defined by $f$.
If we want to consider bigger perturbations, we can work with negative perturbations as the following lemma illustrates.

\begin{lem}\label{lem:neg-pert-imply-f-pos}
    Let $f \in \ZZ[\X]$ and $\lambda > 0$ such that $\negpert{f}(\X) \geq 0$. Then, $f > 0$.
\end{lem}

\begin{proof}
    Observe that $\negpert{f}(\X) \geq 0$ is equivalent to
        $f(\X) 
		\geq  \lambda \sum\nolimits_{i=1}^{n}(1+X_i^2 + X_{i}^{d})$. As $\lambda > 0$, we have that $\lambda \sum\nolimits_{i=1}^{n}(1+X_i^2 + X_{i}^{d}) > 0$, so $f > 0$.
\end{proof}

The problem with negative perturbations is that, even for positive functions, there might not be $\lambda > 0$ such that $\negpert{f} \geq 0$.

\begin{examp}
    Let $f = (X_1 X_2 - \frac{1}{2})^2 + \frac{1}{2}$. This function is positive, but for any $\lambda > 0$, $\negpert{f}(\frac{1}{\sqrt{\lambda}}, 0) < 0$.
\end{examp}

In what follows we restrict ourselves to negative perturbations of coercive functions obtained by the stereographic transformation. The zeros of these functions are related to zeros at infinity.

\begin{lem}
    Let $f \in \ZZ[\X]$ be a polynomial in $n$ variables, of even degree $d$, and $f(\bm{0}) \neq 0$. Then, $\nsteo{f}$ is strictly positive on $\RR^n$ if and only if the highest degree homogeneous part of $f$,  $f^{(d)}$,
    has no zeros on the unit sphere $\Sr^{n-1}$, and $f$ is strictly positive on $\RR^n$.
\end{lem}

\begin{proof}
  By Thm.~\ref{thm:aff-steo-equiv}, the stereographic transformation
  preserves nonnegativity.
  Assume that $\nsteo{f}(\x) = 0$.  By \eqref{eq:steoCases}, either
  $\|\x\|_2 \neq 1$ and there is $\bm{y}\in \RR^n$ such that $f(\bm{y}) = 0$ (as $(-1+\ntwo{\x}^2)^d > 0$), or
  $f^{(d)}(\x) = 0$.
  Alternately, as $f^{(d)}$ is homogeneous, if there is $\x$ such that
  $f^{(d)}(\x) = 0$, we can assume that $\|\x\|_2 = 1$, and so
  $\nsteo{f}(\x) = 0$.
  Assume now that $f(\x) = 0$ for some $\x\in \R^n$. Then $f^h(1,\x)=0$. Scaling down this vector to the unit sphere, we have that $f^h(z_0,\z)=0$ for some $(z_0,\z)\in \mathbb{S}^{n}$ with $z_0\neq 0$. Under stereographic projection, points with zero first coordinates correspond to vectors in $\R^n$ of unit length. It follows that there is $\bm{y}\in \R^n$ such that
  $\|\bm{y}\|_2 \neq 1$ and $\nsteo{f}(\bm{y}) = 0$.
\end{proof}

Like in the case of positive HJ perturbations, in what follows we establish an equivalence of strict positivity between a polynomial and its negative HJ perturbation under the assumption that the highest degree homogeneous part the polynomial has no zeros on the unit sphere.

\begin{thm}\label{thm:minus-sign-perturbation-effective}
    Let $f \in \ZZ[\X]$ be a polynomial in $n$ variables, of even degree $d$, $f(\bm{0}) > 0$, and $H(f) \leq 2^{\tau}$. 
    Let $f^{(d)}$ be the highest degree homogeneous part of $f$ and let $\Rad{n,d,\tau}$ be the constant defined in Eq.~\ref{eq:defOfRad}. For any $\lambda \in \left(0, \frac{1}{6n}\frac{\min \nsteo{f}}{ \Rad{n,d,\tau}^{2d}} \right]$, we have the following equivalence
    \[ \left\{
    \begin{array}{cl}
         f > 0 &  \text{ on }\RR^n, 
     \text{ and }  \\
         f^{(d)}  & \text{ no zeros on } \Sr^{n-1} 
    \end{array} \right\} \iff   
    \negpert{\nsteo{f}} \geq 0 \text{ on }\RR^n .
    \]
    In particular, we can consider any $\lambda \in (0, \epsilon(n,d,\tau)]$, where $\epsilon(n,d,\tau)$ is as defined in  Eq.~\eqref{eq:constantPert}. 
\end{thm}

\begin{proof}
    For the first part of the statement, observe that
    $(\impliedby)$ is implied by Lem.~\ref{lem:neg-pert-imply-f-pos}. Hence we prove the direction $(\implies)$.
    It follows from Eq.~\eqref{eq:steoCases} that, if $d$ is even, $f > 0$ and $f^{(d)} > 0$, then $\nsteo{f} > 0$.
    Moreover, we have that $\min \nsteo{f} \leq \nsteo{f}(\bm{0}) = f(\bm{0}) \leq 2^{\tau}$.
    
    Consider $\lambda \in  \left(0,\frac{1}{2n}\frac{\min \nsteo{f}}{3\Rad{f}^{2d}} \right]$. 
    When $\ntwo{\x} \geq 1 $, the inequality from Eq.~\eqref{eq:ineqSteo}, implies that
    \begin{align*}
    \negpert{\nsteo{f}}(\x)& \geq \nsteo{f}(\x) -  \left(\frac{1}{6n} \frac{\min \nsteo{f}}{\Rad{n,d,\tau}^{2d}}\right) (3 \ n \ \ntwo{\x}^{2d})  \\
     & \geq \ntwo{ \x }^{2d} - 2^d \ {d + n \choose n} \ H(f) \ \ntwo{\x}^{2d - 1} -  \left(\frac{2^{\tau}}{2 \, \Rad{n,d,\tau}^{2d}}\right) \ \ntwo{\x}^{2d}
    \end{align*}
    Hence, $\negpert{\nsteo{f}}$ is coercive and $\negpert{\nsteo{f}} \geq  0$ when
    $\ntwo{\x} \geq 2^{d+\tau+1} \ {d + n \choose n}$, as 
    $2^{\tau} <  \Rad{n,d,\tau}^{2d}$.

    Next, we lower-bound the minimum of $\negpert{\nsteo{f}}$
    on the ball of radius $2^{d+\tau+1} \ {d + n \choose n}$ centered at $\bm{0}$. As the maximum of  $\sum_{i=1}^{n}(1+x_i^2+x_i^{2d})$ on this ball is attained in the boundary, it is upper bounded by $3 n \ (2^{d+\tau+1} \ {d + n \choose n} )^{2d}$.
    Moreover, as $d > 1$, we have that ${d + n \choose n } \leq 2^{d+n-1}$, and so    
    $\Rad{n,d,\tau} \geq 2^{d+\tau+1} \ {d + n \choose n }$.
    Hence, we conclude that $\negpert{\nsteo{f}} \geq 0$, as for any $\ntwo{\x} \leq 2^{d+\tau+1} \ {d + n \choose n}$,
    \begin{align*}
    \negpert{\nsteo{f}}(\x) \geq \min \nsteo{f} -  \frac{1}{6n}\frac{\min \nsteo{f}}{\Rad{n,d,\tau}^{2d}}
    \left(3n \left(2^{d+\tau+1} \ {d + n \choose n }\right)^{2d}\right) 
    = \frac{\min \nsteo{f}}{2}  > 0
    \end{align*}

The second part of the statement follows from Cor.~\ref{cor:boundCritValSteo}, as when $\min \nsteo{f} > 0$,  $\min \nsteo{f} \geq \bd{\nsteo{f}}$.
\end{proof}

In Thm.~\ref{thm:minus-sign-perturbation-effective}, the equivalence
of strict positivity between $f$ and $\negpert{\nsteo{f}}$ depends on
$\lambda$ being small enough.
In practice, if we find $\lambda$ such that $\negpert{\nsteo{f}} > 0$,
then we can use it to certify the positivity of $f$
(Lem.~\ref{lem:neg-pert-imply-f-pos}).
The following lemma shows that, if $\lambda$ is not too big, say
$\lambda < f(\bm{0})$, then $\negpert{\nsteo{f}}$ is still coercive.
Therefore, we can choose any generic-enough $\lambda < f(\bm{0})$ and
verify if $\negpert{\nsteo{f}} > 0$ using \sosrur.
This way, Thm.~\ref{thm:minus-sign-perturbation-effective} gives a
lower bound on how small $\lambda$ can be in the worst-case.
 
\begin{lem}\label{lem:negpertCoerc}
    Let $f \in \ZZ[\X]$ be a polynomial in $n$ variables, of even degree $d$, and $f(\bm{0}) > 0$. If $\lambda < f(\bm{0})$, then $\negpert{\nsteo{f}}$ is coercive.
\end{lem}

\begin{proof}
    By the definition of $\negpert{\nsteo{f}}$, we have
    $\negpert{\nsteo{f}}(\X) \coloneqq \nsteo{f}(\X) 
		- \lambda \sum\nolimits_{i=1}^{n}(1+X_i^2 + X_{i}^{2d})$.
    Note that the highest degree homogeneous part of $\negpert{\nsteo{f}}$ is the degree $2d$ terms in $a_{\bm{0}}(-1 + \ntwo{\X}^2)^d - \lambda\sum_{i=1}^{n}X_i^{2d}$, which is $\sum_{\substack{0 \leq i_1, \dots, i_n < d \\ i_1 + \cdots +i_n = d}} a_{\bm{0}}\binom{d}{i_1, \dots, i_n} X_1^{2i_1}\cdots X_n^{2i_n} + \sum_{i=1}^{n} (a_{\bm{0}} - \lambda) X_i^{2d}$, where $a_{\bm{0}}$ is the constant terms of $f$, i.e $a_{\bm{0}}  = f(\bm{0}) \geq 1$. All these degree $2d$ terms are of positive coefficients and of even powers for each variable $X_1,\dots,X_n$, so $f$ must be coercive.
\end{proof}

\section{Algorithms}
\label{sec:algorithms}

In this section we introduce algorithms to certify that a polynomial is nonnegative, or to prove that it is not. Our algorithms reduce the problem of certification to the setting of Thm.~\ref{thm:SOS-RUR}, that is, certify that a polynomial whose gradient ideal is zero-dimensional and whose minimum is attained, is nonnegative. Using this reduction, we perform our certification via algorithm $\sosrur$.

In most of the cases, the stereographic transformation leads to polynomials satisfying $(\Pi)$, i.e, zero dimensional gradient ideal and minimum attained (by coerciveness, see Lem.~\ref{lem:coercive-has-min}). If this is the case, then we can certify the nonnegativity of $f$ by applying \sosrur to $\nsteo{f}$. 
However, as the next example demonstrates, this is not always the case.

\begin{examp}\label{eg:steo-nonzerodim}
    Let $f(\X) = \ntwo{\X}^2 +4 = X_1^2 + \dots +X_{n}^2+4$. 
    Then, $\nsteo{f}(\X) = 4\ntwo{\X}^2 + 4(-1+\ntwo{\X}^2)^2$. 
    In this case, the gradient variety is
    \[
    	\VV(I_{\nabla \nsteo{f}}) = 
    	\VV(X_1(-1+2\ntwo{\X}^2), \dots, X_n(-1+2\ntwo{\X}^2)) 
    	\supseteq 
    	\VV(-1+2\ntwo{\X}^2).
    \]
    Hence, the gradient ideal of $\nsteo{f}$, $I_{\nabla \nsteo{f}}$, has dimension $n-1$.
\end{examp}

To overcome the obstacle that the gradient ideal of the stereographic transformation of $f$ might not be zero dimensional, 
	we combine the stereographic transformation with the Hanzon-Jibetean perturbations. 
	We introduce two algorithms;
	one based on the positive HJ perturbation and one based on the negative
	HJ perturbation, see Def.~\ref{def:pert-lambda}.
	
	The first one, based on the positive perturbation,
	certifies the nonnegativity of a polynomial $f$,
	by applying the \sosrur algorithm on the polynomial $\pospert{\nsteo{f}}$,
	where $\lambda$ is sufficiently small and depends on the size of $f$.
	We call it \hjpos. It makes no assumption on $f$, other than $f(\bm{0}) > 0$.
	
	The second algorithm, based on the negative perturbation, we call it \hjneg. It certifies the nonnegativity of a polynomial $f$ by 
	applying \sosrur on the polynomial  $\negpert{\nsteo{f}}$.
	Its main difference, and advantage, compared to \hjpos is that 
	$\lambda$ depends on actual minimum of $\nsteo{f}$; in practice
	this is significantly smaller than the $\lambda$ used in \hjpos.
	Even more, we do not need to know the bound a priori.
	However, the algorithm only succeeds to certify the nonnegativity of polynomials
	$f$ when $f$ and its higher degree homogeneous part are positive.
	Nevertheless, we do not need to know if these assumptions are satisfied before running the algorithm.
	If they are not satisfied, then the algorithm reports
	that it cannot handle the input.
	
	Finally, by combining \hjpos and \hjneg we can get the best of the two.
	We introduce the algorithm \hjsosrur, that proceeds as follows:
	First, it checks whether it can decide nonnegativity using only the stereographic transformation.
	If this is not the case, then it applies \hjneg. If the latter fails,
	then it applies \hjpos, which relies on worst case bounds and never fails.
	In this way, either it certifies the nonnegativity of the input polynomial
	or computes one point at which the polynomial has a negative value.

\paragraph{Algorithm \hjpos.} 
The first algorithm relies on the positive HJ perturbation.

\begin{algorithm}[ht]
	\caption{\hjpos}\label{alg:SOS-CS-POS}
	  \textbf{Input:} A nonzero polynomial $f \in \ZZ[\X]$, in $n$ variables, of even degree $d$, such that $H(f)\leq 2^{\tau}$.

   \emph{Assumption:} $f(\bm{0}) > 0$.
    \vspace{2pt}

	\textbf{Output:} 
  	A tuple $[\mathtt{NonNeg}, \mathtt{Cert}]$. 
		\begin{itemize}[leftmargin=1.7cm]
			\item [$\mathtt{NonNeg}$] : a boolean; \texttt{True} if $f$ is nonnegative,  \texttt{False} otherwise.
			\item [$\mathtt{Cert}$] :  a list containing one of the following (depending on the value of $\mathtt{NonNeg}$):
			\begin{itemize}
  				\item If $f$ is nonnegative ($\mathtt{NonNeg} = \texttt{True})$,
  		then  $\mathtt{Cert} = [\lambda, \mathtt{C}]$,  \newline
  		where
        $\lambda \in \Q$ and $\lambda < \epsilon(n,d,\tau)$ (see Thm.~\ref{thm:f-pos-effect-equiv-pert-steo}), $\pospert{\nsteo{f}}$ satisfies $(\Pi)$, and \newline
        $\mathtt{C}$ is a valid certificate of nonnegativity for $\pospert{\nsteo{f}}$.
        \item If $f$ is not nonnegative ($\mathtt{NonNeg} = \texttt{False}$), then $\mathtt{Cert} = [\x]$, \newline
       	where $\x \in \Q^n$ is such that $f(\x) < 0$.
		\end{itemize}
	\end{itemize}

\textbf{Steps:}
    \begin{enumerate}
		\item Let $\epsilon(n,d,\tau)$  be the constant
        of Eq.~\eqref{eq:constantPert}, that is, \hfill (see Thm.~\ref{thm:f-pos-effect-equiv-pert-steo})
        \[\epsilon(n,d,\tau) \coloneqq \frac{1}{6n}\frac{\bd{n,  2d, d\lg(n+1)+ n+2d+\tau}}{\Rad{n,d,\tau}^{2d+2}}.\]
        \item Choose any $\lambda \in (0, \epsilon(n,d,\tau)]$ \ \ \ (the simplest rational in the interval) \hfill (See Rmk.~\ref{rmk:bitsizePosPert})
		\item Compute $\pospert{\nsteo{f}} = \nsteo{f} + \lambda \sum_{i=1}^{n} (1+ X_i^2 + X_i^{2d+2})$ \hfill (Def.~\ref{def:pert-lambda})
	
		\item $[\mathtt{NonNeg}, \mathtt{C}] \leftarrow \sosrur(\pospert{\nsteo{f}})$
		 
		\item If $\mathtt{NonNeg} = \mathtt{True}$ then  \\ ${ } \qquad\textsc{return }\  [\mathtt{True}, [\lambda, \mathtt{C}]]$
		$\qquad\quad $ // $f$ is nonnegative, return the certificate
		\item If $\mathtt{NonNeg} = \mathtt{False}$ then 
		 \hspace{1.5cm} // in this case $\texttt{C} = [\bm{p}]$ \\
                ${ } \qquad \textsc{return }\   [\mathtt{False}, [\mathscr{Q}(f,\bm{p})]]$ $\quad$  // return the witness point \hfill (See Lem.~\ref{lem:stereo-witness})    
    \end{enumerate}

\end{algorithm}

\begin{thm} \label{correctAlgPos}
    Algorithm {\rm \hjpos} (Alg.~\ref{alg:SOS-CS}) is correct. 
    In particular, 
    \begin{itemize}
    \item If $f$ is nonnegative, then it outputs $\lambda \in \Q$, such that $\pospert{\nsteo{f}}$ satisfies $(\Pi)$ and a valid certificate of nonnegativity for $\pospert{\nsteo{f}}$. The existence of this certificate guarantees that $f$ is nonnegative. 

   \item If $f$ is not nonnegative, then it outputs a (witness) point $\bm{x} \in \Q^n$, such that $f(\bm{x}) < 0$. 
   \end{itemize}
\end{thm}
\begin{proof}
    At line (2), we choose a $\lambda  \in (0, \epsilon(n,d,\tau)]$.
    By Thm.~\ref{thm:f-pos-effect-equiv-pert-steo}, for any such $\lambda$,
     $f$ is nonnegative if and only if
    $\pospert{\nsteo{f}}$ is nonnegative. Moreover, $\pospert{\nsteo{f}}$ satisfies $(\Pi)$.
    Therefore, the output of \sosrur with input $\pospert{\nsteo{f}}$
    is,  either a valid certificate of nonnegativity for $\pospert{\nsteo{f}}$, or a point $\bm{p} \in \Q^n$, such that $\pospert{\nsteo{f}}(\bm{p}) < 0$ \cite{btz-sosrur-hal-25}.
    In the second case, we have that $0 > \pospert{\nsteo{f}}(\bm{p}) >  \nsteo{f}(\bm{p})$, so
    by Lem.~\ref{lem:stereo-witness}, we have that $f(\mathscr{Q}(f,\bm{p})) < 0$ and $\mathscr{Q}(f,\bm{p}) \in \Q^n$.
\end{proof}

\begin{rmk}[Verification of the certificate]
\label{rem:verify-hjpos}
    Consider $f \in \Q[\x]$ of degree $d$ such that $H(f)\leq 2^{\tau}$ and $f(\bm{0}) > 0$.
    If \hjpos asserts that $f$ is nonnegative, then it outputs $[\mathtt{True}, [\lambda,\mathtt{Cert}]]$.
    We verify the consistency of the output in two steps.
    First, we check whether $\lambda < \epsilon(n,d,\tau)$. If this is the case, then $\pospert{\nsteo{f}}$ satisfies $(\Pi)$ (Def.~\ref{def:pi} and Thm.~\ref{thm:f-pos-effect-equiv-pert-steo}).
    The second step is to verify that $\mathtt{Cert}$ is indeed a valid certificate of nonnegativity for $\pospert{\nsteo{f}}$;
    for this we use the algorithm supported by Thm.~\ref{thm:verifyCertIfPi}.
\end{rmk}

\begin{rmk}
    [About the bitsize of $\lambda$]
    \label{rmk:bitsizePosPert}
    As $\lambda \in (0, \epsilon(n,d,\tau)]$, we can assume that the
    worst case bitsize of $\lambda$ is
    $\bsz(\lambda) = \bsz (\epsilon(n,d,\tau)) = \OO(n \, 2^n \, d^n
    \,({ n^2 d} + \tau))$; see Eq.\eqref{eq:constantPert}.  In practice,
    we might choose the simplest rational in this interval by
    considering the continued fraction expansion of
    $\epsilon(n,d,\tau)$.
\end{rmk}

\paragraph{Algorithm \hjneg.}
The second algorithm relies on the negative HJ perturbation (Def.~\ref{def:pert-lambda}). It does not work with rationals having the worst case bitsize
right from the beginning, as \hjpos does, and hence it is more efficient in practice.
On the negative side, it requires that both $f$ and $f^{(d)}$, that is
the highest degree homogeneous part of $f$, are positive, so the
algorithm can fail.

\begin{algorithm}[ht]
	\caption{\hjneg }\label{alg:SOS-CS-NEG}
    
	  \textbf{Input:} A nonzero polynomial $f \in \ZZ[\X]$, in $n$ variables, of even degree $d$, such that $H(f)\leq 2^{\tau}$.

   \emph{Assumption:} $f(\bm{0}) > 0$.
    \vspace{2pt}

	\textbf{Output:} 
		One of the following: 
		\begin{itemize}[leftmargin=2.5cm]
			\item [{$[\lambda, \mathtt{Cert}]$}  : ] 
			if $f$ \emph{and} its highest degree homogeneous part $f^{(d)}$ are  positive. \newline
			We have that $\lambda \in \Q$, $\negpert{\nsteo{f}}$ satisfies $(\Pi)$ 
			and \newline $\mathtt{Cert}$ is a valid certificate of nonnegativity of $\negpert{\nsteo{f}}$.
			\item [$\mathtt{Fail}$ : ]   if $f$ \emph{or} $f^{(d)}$ are not positive.
		\end{itemize}

\textbf{Steps:}
    \begin{enumerate}
        \item Let $\gamma \leftarrow \frac{1}{2}$ and compute the constant $\eps := \epsilon(n,d,\tau)$, as in Eq.~\eqref{eq:constantPert} \hfill (see Thm.~\ref{thm:f-pos-effect-equiv-pert-steo})
        \item
	    While $2^{-\gamma} > \eps$
        \begin{enumerate}        
		\item $\gamma \leftarrow 2 \ \gamma$
        \item Choose $\lambda \in (2^{- 2 \ \gamma}, 2^{-\gamma}]$. \hfill (See Rmk.~\ref{rmk:unlucky})
        \item If the gradient ideal of $\negpert{\nsteo{f}}$ is not zero-dimensional 
        or  has solutions ``at infinity'', \newline Go to (b). 
       
        \item Compute $\negpert{\nsteo{f}} = \nsteo{f} - \lambda \sum_{i=1}^{n} (1+ X_i^2 + X_i^{2d})$ \hfill (Def.~\ref{def:pert-lambda})
            		\item $[\mathtt{NonNeg}, \mathtt{Cert}] \leftarrow \sosrur(\negpert{\nsteo{f}})$ 
        \item If $\mathtt{NonNeg} = \mathtt{True}$ then  \\
        	${ } \qquad\textsc{return }\  [\lambda, \mathtt{Cert}]$
        \end{enumerate}

        \item \textsc{return } $\texttt{Fail}$
    \end{enumerate}

\end{algorithm}

\begin{thm}\label{thm:correctAlgMinus}
    Algorithm \hjneg  (Alg.~\ref{alg:SOS-CS-NEG}) is correct. In particular,
    \begin{itemize}
        \item   If $f$  and its highest degree homogeneous part are positive, then it returns $\lambda \in \Q$ and 
        $\mathtt{Cert}$ such that  $\negpert{\nsteo{f}}$ satisfies condition $(\Pi)$ and $\mathtt{Cert}$ is a valid certificate of nonnegativity for $\negpert{\nsteo{f}}$. Moreover, the existence of this certificate implies that $f$ is positive.
        \item If the algorithm fails, then either $f$ or its highest degree homogeneous part $f^{(d)}$ are not positive.
    \end{itemize}
\end{thm}

\begin{proof}
	Starting from $\gamma = 1$,
	\hjneg repeatedly doubles $\gamma$ until, 
	either $2^{-\gamma} < \epsilon(n,d,\tau)$
	or we can decide the nonnegativity (or not) of  $\negpert{\nsteo{f}}$ 
	using \sosrur.
	The bound on $\gamma$ guarantees termination; provided
	that we loop a finite number of times between steps (b) and (c).

    As $\lambda < 1 \leq f(\bm{0})$, 
    Lem.~\ref{lem:negpertCoerc} implies that $\negpert{\nsteo{f}}$ is coercive
    and so it attains its infimum.
    Moreover, if $\lambda$ is generic, then 
    the gradient ideal of $\negpert{\nsteo{f}}$ is zero dimensional (and has not solutions at infinity), by Prop.~\ref{prop:0-dim-rad-pert-generically};
    hence it satisfies condition $(\Pi)$.
    Lem.~\ref{lem:unluckyChoicesNegPert} implies that there are at most  $(2d - 1)^{n-1} (2d - 1 + n)$ values for $\lambda \in \Q$, such that $\negpert{\nsteo{f}}$ does not satisfy condition $(\Pi)$ (or has solutions at infinity). So, for every $\gamma$,
    we loop a finite number of times between steps (b) and (c).
    This establishes the termination of the algorithm.

    Moreover, if we choose \emph{any} $\lambda < \epsilon(n,d,\tau)$, 
    by Thm.~\ref{thm:minus-sign-perturbation-effective}, then
    $\negpert{\nsteo{f}}$ is nonnegative
    if and only if both $f$ and $f^{(d)}$ are positive.
    Thus, if both $f$ and $f^{(d)}$ are positive, 
    any $\lambda$ computed from the loop of steps (b) and (c)
    enforces 
    $\negpert{\nsteo{f}}$ to satisfy condition $(\Pi)$,
    and thus, 
    we can certify the  nonnegativity of $\negpert{\nsteo{f}}$ using \sosrur.
 
    On the other hand, if $f$ or $f^{(d)}$ are not positive, we cannot certify the positivity of $f$ from the nonnegativity of $\negpert{\nsteo{f}}$.
    The algorithm exhausts all the values of $\gamma$ and outputs fail. 
\end{proof}

\begin{rmk}[Lucky choices and bitsize of $\lambda$] \label{rmk:unlucky}
    Notice that, if we choose a \emph{generic} $\lambda$ at the step (b) of the algorithm, 
   	 then, by Prop.~\ref{prop:0-dim-rad-pert-generically}, 
   	 $\negpert{\nsteo{f}}$ will satisfy $(\Pi)$ and its gradient ideal will have no solutions ``at infinity''.
         Hence, there are only finite \emph{unlucky} $\lambda$'s that
         do not satisfy the previous properties and so, they do not go
         through step (c). These $\lambda$'s are characterised by
         Lem.~\ref{lem:unluckyChoicesNegPert}.
         If we combine this lemma with Schwartz-Zippel lemma,
         e.g.~\cite{lipton}, then we can lower bound the probability
         of finding a \emph{lucky} $\lambda$.  Fix $k \in \N$ and
         consider a set $S \subset \Q$ with
         $k \cdot (2d - 1)^{n-1} (2d - 1 + n)$ different elements. The
         probability that a uniformly random chosen $\lambda \in S$ is
         \emph{unlucky} is upper-bounded by $\frac{1}{k}$.
         Moreover, given $\gamma \in \N$, we can choose a set
         $S \subset (2^{-2 \gamma}, 2^{-\gamma}]$ such that its
         elements have bitsize at most
         $\gamma + 2 + \lg(k \cdot (2d - 1)^{n-1} (2d - 1 + n)) =
         \sO(\gamma + n \lg d + \lg k).$
    For this reason, we say that $\lambda  \in (2^{-2 \gamma}, 2^{-\gamma}]$ has \emph{small bitsize} when its bitsize is $\sO(\gamma + n \lg d + \lg k)$.
\end{rmk}

\begin{rmk}[Verification of the certificate]
\label{rem:verify-hjneg}
    To verify that the algorithm does not fail when $f$ is positive,
     we first need to check that $0 < \lambda < f(\bm{0})$. In this case, $\negpert{\nsteo{f}}$ coercive, and so its minimum is attained.
    Therefore, we can verify if $\mathtt{Cert}$ is a valid certificate of nonnegativity for $\negpert{\nsteo{f}}$  using Thm.~\ref{thm:verifyCertIfPi}. 
\end{rmk}

\paragraph{Algorithm \hjsosrur.} 
	We combine the algorithms \hjpos and \hjneg to introduce an algorithm
	that have the advantages of both. We call the algorithm \hjsosrur to emphasize that is based on (both) HJ perturbations (Def.~\ref{def:pert-lambda}) and \sosrur \cite{btz-sosrur-hal-25}.

\begin{algorithm}[h]
	\caption{\hjsosrur}\label{alg:SOS-CS}\label{alg:hjsosrur}
	
   \textbf{Input:} A nonzero polynomial $f \in \ZZ[\X]$, in $n$ variables of even degree $d$, such that $H(f)\leq 2^{\tau}$.

   \emph{Assumption:} $f(\bm{0}) > 0$.
    \vspace{2pt}

	\textbf{Output:} 
  	A tuple $[\mathtt{NonNeg}, \mathtt{PertType}, \mathtt{Cert}]$. 
	\begin{itemize}[leftmargin=2.cm]
		\item [{$\mathtt{NonNeg}$}: ]
		a Boolean; \texttt{True} if $f$ is nonnegative,  \texttt{False} otherwise.
		\item [$\mathtt{PertType}$: ] 
		an element of $\{\texttt{NoPert},\texttt{NegPert}, \texttt{PosPert}\}$ indicating whether we applied no, a negative, or a positive HJ perturbation to $f$; see Def.~\ref{def:pert-lambda}.
		\item [$\mathtt{Cert}$:] a list, the contents of which depend on the other two outputs. In particular: 
		\begin{itemize}[leftmargin=.5cm]
			\item If $\mathtt{NonNeg} = \texttt{False}$, then 
			$\mathtt{Cert} = [ \x ]$, where  $\x \in \Q^n$
			is such that  $f(\x) < 0$.
			\item If $\mathtt{NonNeg} = \texttt{True}$, then
			\begin{itemize}
                        \item If
                          $\mathtt{PertType} = \texttt{NoPert}$, then
                          $\nsteo{f}$ satisfies condition $(\Pi)$
                          (Def.~\ref{def:pi}) and \newline
                          $\texttt{Cert}$ is a valid certificate of
                          nonnegativity for $\nsteo(f)$.
				\item If $\mathtt{PertType} = \texttt{NegPert}$, 
				then $\mathtt{Cert} = [\lambda, \mathtt{C}]$, \newline
				where $\lambda \in \Q$ and $\lambda < f(\bm{0})$, $\negpert{\nsteo{f}}$ satisfies $(\Pi)$,
				and \newline
				$\mathtt{C}$ is a valid certificate of nonnegativity for $\negpert{\nsteo{f}}$. 
				In this case, $f$ is positive. 
				\item If $\mathtt{PertType} = \texttt{PosPert}$,
				then $\mathtt{Cert} = [\lambda, \mathtt{C}]$, \newline
				where $\lambda \in \Q$ and $\lambda < \epsilon(n,d,\tau)$ (see Thm.~\ref{thm:f-pos-effect-equiv-pert-steo}), $\pospert{\nsteo{f}}$ satisfies $(\Pi)$, and \newline 
				$\mathtt{C}$ is a valid certificate of nonnegativity for $\pospert{\nsteo{f}}$.
			\end{itemize} 
		\end{itemize}
	\end{itemize}
    
\textbf{Steps:}

    \begin{enumerate}
    \item If $I_{\nabla \nsteo{f}}$ is zero-dimensional and has no solutions ``at infinity'', then
    \begin{enumerate}
        \item  $[\mathtt{NonNeg}, \mathtt{Cert}] \leftarrow \sosrur({\nsteo{f}})$
        \item If $\mathtt{NonNeg} = \mathtt{True}$, \! \quad\quad $\textsc{return }\ [\mathtt{True},\mathtt{NoPert}, \mathtt{Cert}]$
        \item Else, $\mathtt{Cert} = [\bm{p}]$ \;\;\;\qquad $ \textsc{return }\ [\mathtt{False}, - , [\mathscr{Q}(f,\bm{p})]]$ \hfill (See Lem.~\ref{lem:stereo-witness})
    \end{enumerate}

    \item Else, try $[\lambda, \mathtt{Cert}] \leftarrow \hjneg(f)$
    \begin{enumerate}
    \item If does not $\texttt{Fail}$, \qquad\; $ \textsc{return }\ [\mathtt{True},\mathtt{NegPert}, [\lambda, \mathtt{Cert}]]$

    \item Else, $[\mathtt{NonNeg}, \mathtt{Cert}] \leftarrow  \hjpos(f)$
    \begin{enumerate}
    \item If $\mathtt{NonNeg} = \mathtt{True}$, \!\,  $ \textsc{return }\ [\mathtt{True},\mathtt{PosPert}, \mathtt{Cert}]$
    \item Else,  \!\;\qquad\qquad\qquad $ \textsc{return }\ [\mathtt{False},-, \mathtt{Cert}]$
    \end{enumerate}
    \end{enumerate}
    \end{enumerate}
\end{algorithm}

\begin{thm}
    Algorithm \hjsosrur (Alg.~\ref{alg:hjsosrur}) is correct, that is,
    \begin{itemize}
        \item If $f$ is not nonnegative, then it outputs an $\bm{x} \in \Q^n$, such that $f(\bm{x}) < 0$.
        \item Otherwise, the algorithm returns polynomials whose
          nonnegativity imply the nonnegativity of $f$ together with
          valid certificates of nonnegativity for them (see the output
          of \hjsosrur in Alg.~\ref{alg:hjsosrur}).
    \end{itemize}
\end{thm}

\begin{proof}
    Assume first that $I_{\nabla \nsteo{f}}$ is zero-dimensional. As $\nsteo{f}$ is coercive, it attains its minimum. Hence, $f$ satisfies condition $(\Pi)$. In this case, we can certify if $\nsteo{f}$ nonnegative using \sosrur. If this is the case, then, by Thm.~\ref{thm:aff-steo-equiv}, a valid certificate of nonnegativity for $\nsteo{f}$ implies that $f$ is nonnegative. If $f$ is not nonnegative, then we will obtain a point $\bm{x} \in \Q$ such that $\nsteo{f}(\bm{x}) < 0$ and so, by Lem.~\ref{lem:stereo-witness}, $\mathscr{Q}(f,\bm{p}) \in \Q^n$ and $f(\mathscr{Q}(f,\bm{p})) < 0$.

    If we are not sure if $I_{\nabla \nsteo{f}}$ is zero-dimensional, then we try to certify that $f$ is positive using \texttt{SOS-CS-NEG}. By Thm.~\ref{thm:correctAlgMinus}, if the algorithm does not fail, $\negpert{\nsteo{f}}$ satisfies $(\Pi)$, and       
        $\mathtt{C}$ is a valid certificate of nonnegativity for $\negpert{\nsteo{f}}$. By Lem.~\ref{lem:neg-pert-imply-f-pos}, this certificate implies that $f$ is \emph{positive}.

    Finally, if we could not certify the nonnegativity of $f$ using the previous approaches, we use  \texttt{SOS-CS-POS}, which by Thm.~\ref{correctAlgPos}, leads either to a certificate of nonnegativity for $f$ or to a rational point where the function vanishes.
\end{proof}

\begin{rmk}[Verification]
\label{rem:verify-hjsosrur}
    Remarks~\ref{rem:verify-hjpos} and \ref{rem:verify-hjneg} 
    detail how to verify the certificates induced by \hjpos and \hjneg, respectively. 
    Hence, to verify the certificate induced by \hjsosrur,
    it suffices to verify the certificate we obtain when $\mathtt{PertType} = {\mathtt{NoPert}}$. In this case, we certify the nonnegativity of
     $\nsteo{f}$. As $\nsteo{f}$ is coercive, it attains its infimum,
     so the validity of the certificate follows from Thm.~\ref{thm:verifyCertIfPi}. 
\end{rmk}

\paragraph{On the assumption $f(\bm{0}) > 0$.}
The algorithms that we present assume that the input polynomial $f$ has a positive constant coefficient.
If this coefficient is zero, then 
$\nsteo{f}$ may not be a coercive function, and so our perturbation schemes, Thm.~\ref{thm:f-pos-equiv-pertf-pos}, do not hold.
To ensure this property, we need to translate $f$. 
For this, we find a $\bm{c} = (c_1, \dots, c_n) \in \mathbb R^n$,
 such that $f(\bm{c}) \neq 0$.
 Then, we apply \hjsosrur on $f(X_1 + c_1, \dots,X_n + c_n)$, if $f(\bm{c}) > 0$; otherwise we return the point $\bm{c}$. 
 We can find a suitable $\bm{c}$ using a randomized procedure
 that has probability of success arbitrarily close to 1,
 by applying straightforwardly Schwartz-Zippel lemma, 
 see \cite{lipton} and references therein.
 
\begin{lem}\label{lem:translation-of-variables}
    Let $f$ be a polynomial of degree $d$ and fix a constant $k \in \mathbb N$. Let $S$ be the grid of points in $\Z^n$ whose coordinates belong to the set $\{1,\dots,k \cdot d\}$.
    If we choose $\bm{c} \in S$ uniformly at random, 
    then the  probability that $f(\bm{c}) = 0$ is at most $\frac{1}{k}$.
\end{lem}

When we translate $f$, we also increase its bitsize; this (slightly) affects
the complexity of \hjsosrur.
We address this issue in Lem.~\ref{lem:translation-of-variables-bitsize}.

\section{Complexity analysis} \label{sec:complexity}

We study the complexity of Alg.~\hjsosrur and its subroutines. 
Our main ingredient is the bit complexity of the \sosrur (Alg.~\ref{alg:SOS-RUR}). Because this is a Monte Carlo algorithm, our algorithms will be of the same kind, and their the probability of success will be the same as \sosrur; see Thm.~\ref{thm:sos-rur-complexity}.

We start this section by studying
the effect of the Hanzon-Jibetean perturbations
on the (bit)size of the polynomials.

\begin{lem} \label{thm:pert-coef-size}
    Let $f \in \ZZ[\X]$
    be a polynomial in $n$ variables of degree $d$ such that $H(f) \leq 2^{\tau}$.
    Let $\lambda \in \Q$, such that $\bsz(\lambda) = \bsz(\epsilon(n,d,\tau))$ (see Eq.~\eqref{eq:constantPert} and Thm.~\ref{thm:minus-sign-perturbation-effective}), then
    \begin{align*}
        \bsz(\pospert{\nsteo{f}}), \bsz(\negpert{\nsteo{f}}), \bsz(\lambda) 
        = \sO(n \, 2^n \, d^n \, ({ n^2  d} + \tau)).
    \end{align*}
\end{lem}
\begin{proof}
	The polynomials $\pospert{\nsteo{f}}$ and $\negpert{\nsteo{f}}$
	have similar bitsize, so it suffices to consider one of them. 
    By Lem.~\ref{thm:steo-coef-size},
    $\bsz(\nsteo{f}) = \OO(d\lg(n+1)+ n+2d+\tau) = \sO(n + d + \tau)$.
   	Hence, the bitsize of the $\lambda$ dominates the bound.
   	As $\bsz(\lambda) = \bsz(\epsilon(n,d,\tau)) = 
\sO(n \, 2^n \, d^n \, ({ n^2  d} + \tau))$, the bound follows.
\end{proof} 

Next, we study the complexity of Algorithm {\hjpos}. 

\begin{thm}[Complexity of \hjpos]
	\label{thm:sos-cs-pos-complexity}
	Consider $f \in \ZZ[\X]$ of size $(d, \tau)$ and even degree. 
    Algorithm {\hjpos} (Alg.~\ref{alg:SOS-CS-POS})  computes
    \begin{itemize}
        \item If $f$ is nonnegative, a constant $\lambda \in \Q$ of bitsize
        $\sO(2^n \, d^n \, (d + \tau))$ and a valid certificate of nonnegativity of $\pospert{\nsteo{f}}$ of the total bit-size $\sO(2^{4n} (d+1)^{4n+2} \ ( d + \tau))))$.
        \item If $f$ is not nonnegative, a rational point $\bm{p} \in \Q^n$ such that $f(\bm{p}) < 0$. The bitsize of the witness point is $\sO(2^{3n} (d+1)^{3n + 3}   ( d + \tau))$.
    \end{itemize}
 	The bit complexity of \hjsosrur is
    $\sOB( e^{2 (\omega +1)n} \, (d+1)^{(\omega +2)n}\,  \tau)$, 
	where $\omega$ is the exponent of matrix multiplication. 	
\end{thm}

\begin{proof}
    The complexity bound follows from Thm.~\ref{thm:sos-rur-complexity}, taking into account that, by Thm.~\ref{thm:0-dim-plus-pert}, the gradient ideal  of $\pospert{\nsteo{f}}$ has no solutions ``at infinity''.
    Observe that, by Lem.~\ref{thm:pert-coef-size}, we have that the bitsize of $\pospert{\nsteo{f}}$ and $\lambda$ is $\sO(2^n \, d^n \, (d + \tau))$.
    Moreover, by Lem.~\ref{lem:stereo-witness}, for any $\bm{x} \in \Q^n$, the bitsize of $\mathscr{Q}(f,\x)$ is $\OO(n \, d \, (\lg H(\x)) + \bsz(f))$.
\end{proof}

In what follows, we study the expect and worst-case complexity of \hjneg. 

\begin{thm}[Complexity of \hjneg]
	\label{thm:sos-cs-neg-complexity}
	Consider $f \in \ZZ[\X]$ of size $(d, \tau)$ and even degree. Let $f^{(d)}$ be the highest degree homogeneous part of $f$. Algorithm {\hjneg } (Alg.~\ref{alg:SOS-CS-NEG}) is randomized\footnote{If \sosrur was deterministic or Las Vegas, then \hjneg would be Las Vegas.} and its expected bit complexity is as follows.
    \begin{itemize}
        \item If $f$ and $f^{(d)}$ are positive, the algorithm computes $\lambda \in \Q$ of bitsize $\sO\left( d \ ( d + n + \tau) + |\lg(\min \nsteo{f})|\right)$
         and a valid certificate of nonnegativity for $\negpert{\nsteo{f}}$ of total bitsize \linebreak $\sO(2^{3n} d^{3n + 2} (d^2 + d \ \tau + |\lg(\min \nsteo{f})| ))$ in expected $\sOB(e^{(2 \omega+1) n} d^{(\omega +1)n} \ (d +  \tau + |\lg(\min \nsteo{f})|))$ bits operations. 
         \item Otherwise, the algorithm fails in expected $\sOB(e^{2(\omega+1) n} \ d^{(\omega +2)n} \ \tau )$ bit operations.
    \end{itemize}
    The worst-case complexity of the algorithm is  $\sOB(e^{(2\omega+3) n} \ d^{(\omega +3)n} \ \tau )$.
\end{thm}

\begin{proof}
    We want to estimate the expected complexity of our algorithm.
    To do so, we define $\gamma^{\max}$ as the biggest $\gamma$ that
    our algorithm considers. As we do exponential search, it follows from
    Thm.~\ref{thm:minus-sign-perturbation-effective} that, if $f$ and $f^{(d)}$ are positive, we can assume $\gamma^{\max} = 
    \OO \left(\lg \left(\Rad{n,d,\tau}^{2d} \right) + |\lg(\min \nsteo{f})| \right)$ or, if this is not the case,  
    $\gamma^{\max} = 
    \OO(|\lg(\epsilon(n,d,\tau)|)$. 
    Observe that we will go through the loop of step 2 $\OO(\lg(\gamma^{\max}))$ times.
    
    Recall that the complexity of checking if the gradient ideal of $\nsteo{f}$ is zero-dimensional and has no solutions ``at infinity'' (step (c)) is upper-bounded by the complexity of computing $\sosrur(\negpert{\nsteo{f}})$ (step (e)); see Rmk.~\ref{rmk:verify0dim}.   
    Moreover, this complexity increases with the bitsize of $\lambda$, which is at most $\sOB(\gamma^{\max} + n \lg d + \lg k)$;  where $k$ is as defined in Rmk.~\ref{rmk:unlucky}.
    Hence, by Thm.~\ref{thm:sos-rur-complexity}, the complexity of computing $\sosrur(\negpert{\nsteo{f}})$ each time is at most
    \begin{align}
            \label{eq:complexityNegPert}
    \sOB(e^{(2\omega+1) n} d^{(\omega +1)n-1} (d + \gamma^{\max}  + \tau + \lg k)) .
    \end{align}
        
    Finally, observe that the probability of choosing an \emph{unlucky} $\lambda$, such that we loop in step (c), is upper-bounded by $\frac{1}{k}$. So for each value of $\gamma$, we expect to iterate at most $\frac{k}{k-1}$ times in this step.
    Therefore, the expected total complexity of the algorithm is as in Eq.~\ref{eq:complexityNegPert}.
    If either $f$ or $f^{(d)}$ are not positive, then the algorithm fails after performing (expected)
    $\sOB( e^{2(\omega+1) n} \ d^{(\omega +2)n} \ (\tau + \lg k))$ bit operations.
    If both  $f$ and $f^{(d)}$ are positive, then the algorithm certifies the nonnegativity of  
     $\negpert{\nsteo{f}}$ in an expected complexity of the algorithm is 
    $\sOB(e^{(2 \omega+1) n} d^{(\omega +1)n} \ (d +  \tau + |\lg(\min \nsteo{f})| + \lg k))$.
    In this case, $\lambda$ has bitsize $\sO\left( d \ ( d + n + \tau) + |\lg(\min \nsteo{f})| +\lg k\right)$ and the bitsize of the certificate of nonegativity for  $\negpert{\nsteo{f}}$ is
    $\sO(2^{3n} d^{3n + 2} (d^2 + d \ \tau + |\lg(\min \nsteo{f})| +\lg k))$.
    
    In the worse case, that is, where our algorithm chooses every possible \emph{unlucky} $\lambda$, the complexity of our algorithm is $\sOB(
    e^{(2\omega+3) n} \ d^{(\omega +3)n} \ (\tau + \lg k) )$, by Rmk.~\ref{rmk:unlucky}.
\end{proof}

\begin{thm}
	\label{thm:sos-cs-complexity}
    Consider $f \in \ZZ[\X]$ of size $(d, \tau)$ and even degree. 
    Algorithm {\hjsosrur } (Alg.~\ref{alg:SOS-CS}) is randomized\footnote{As \hjneg, if \sosrur was deterministic or Las Vegas, then \hjsosrur would be Las Vegas.} and its expected complexity is $\sOB( e^{2 (\omega +1)n} \, (d+1)^{(\omega +2)n}\,  \tau))$ bit operations. In the worst-case, the bit complexity is $\sOB( e^{(2\omega+3)n} \, d^{(\omega +3)n}\,  \tau))$.
    
    The size of the output is as follows.
    \begin{itemize}
        \item If $f$ is nonnegative, the algorithm outputs a rational polynomial of bitsize
        $\sO(2^n \, d^n \, (d + \tau))$ and a valid certificate of nonnegativity for it of total bit-size $\sO(2^{4n} (d+1)^{4n+2} \, ( d + \tau))))$ (see output of \hjsosrur in Alg.~\ref{alg:hjsosrur}).
        \item If $f$  is not nonnegative, a rational point $\bm{p} \in \Q^n$ such that $f(\bm{p}) < 0$. The bitsize of the witness point is  $\sO(2^{3n} (d+1)^{3n + 3}  ( d + \tau))$.
    \end{itemize}
\end{thm}
    
\begin{proof}
    In the average case, the complexity of the algorithm is dominated by the complexity of \hjpos (Thm.~\ref{thm:sos-cs-pos-complexity}).
    In the worst case, the complexity is dominated by the worst-case complexity of \hjneg (Thm.~\ref{thm:sos-cs-neg-complexity}).
\end{proof}

\paragraph{On the assumptions $f(\bm{0}) > 0$.}
Given $k \in \N$ and a point $(c_1,\dots,c_n) \in \{1,\dots,k \cdot
d\}^n$, our next lemma offers an upper bound for the bitsize of the
polynomial $f(X_1 + c_1, \dots,X_n + c_n)$.

\begin{lem}\label{lem:translation-of-variables-bitsize}
    Let $f$ be a polynomial of degree $d$ and fix a constant $k \in \mathbb N$. Let $S$ be the grid of points in $\Z^n$ whose coordinates belong to the set $\{1,\dots,k \cdot d\}$.
    If $\bsz(f) \leq \tau$, then $\bsz(f(X_1 + c_1, \dots,X_n + c_n)) = \OO(\tau + n \ d \ (\lg (k) + \lg(d)))$. 
\end{lem}

\begin{proof}
    First, observe that $\bsz(c_i) = \OO(\lg k + \lg d)$. Second, note
    that if $f$ is an univariate polynomial, then $\bsz(f(X_1 + c_1)) \leq \tau + \lg(d+1) + \lg d + d \lg c_1$. Hence, $\bsz(f(X_1 + c_1,X_2, \dots,X_n)) = \OO(\tau + d \lg (k) + d \lg(d))$. Applying the previous equality for each variable $X_i$, we conclude that $\OO(\tau + n \ d \ (\lg (k) + \lg(d)))$.
\end{proof}

\section{SOS perturbations}
\label{sec:sos-perturbations}

Sums of squares polynomials form an important subclass of nonnegative polynomials that nowadays plays a key role in optimization of real polynomials over semi-algebraic sets, see \cite{parrilo_structured_2000} and \cite{lasserre_global_2001}. They form a closed convex full-dimensional subcone of the cone of all nonnegative polynomials of a given degree. While not every nonnegative polynomial is a sum of squares, Artin's solution \cite{artin_uber_1927} of the Hilbert's $17$th problem shows that every nonnegative polynomial becomes a sum of squares after multiplication by the square of some real polynomial.
Checking whether a given polynomial is a sum of squares boils down to verifying whether a certain affine-linear subspace associated to the polynomial non-trivially intersects the cone of positive semi-definite matrices.
Because of their importance both for theory and applications, there has been a lot of interest in different aspects of sums of squares polynomials and their relation to general nonnegative polynomials, see \cite{blekherman_there_2006}, \cite{Uniform}, \cite{lasserre_global_2001} and references therein.
	Lasserre~\cite{l-sos-approx-06} proved that for a nonnegative polynomial $f\in \R[\X]$ and any $\varepsilon>0$, for all sufficiently large $t\in \mathbb{N}$ the perturbed polynomial $f+\varepsilon \theta_t$ is a sum of squares, where $\theta_t=\sum_{k=0}^t \sum_{i=1}^n \frac{X_i^{2k}}{k!}$. Here we study a similar but different perturbation,
\begin{align}\label{eq:fte}
    f_{t,\varepsilon}\ :=\ f+\varepsilon (1+\Vert \X\Vert_2^2)^t,\quad \Vert \X\Vert_2^2\ =\ \sum_{i=1}^n X_i^2.
\end{align}
As we remark in Subsection \ref{sub:comparison}, it follows from Lasserre's result~\cite{l-sos-approx-06} that $f_{t,\varepsilon}$ is a sum of squares for all sufficiently large $t\in \mathbb{N}$. We now prove this fact in a completely different way giving at the same time bounds on $t$.

\begin{thm}
	\label{thm:sos-under-pert}
Let $f = \sum_{|\alpha| \leq d} f_{\alpha} \X^{\alpha} \in \RR[\X]$ 
be a nonnegative polynomial of degree $2d$ and let $\eps > 0$.
Then, for any $t \in \NN$, such that
\[
t  \geq \max \left\{ d,  
\left\lceil 
\frac{1}{\eps} \left( \none{f} +  \frac{\ntwo{\nabla{f}(\bm{0})}^2 }{f_0+ \eps} \right) 
\right\rceil 
\right\}, 
\]
the degree $2t$ polynomial $\fte(\X) = f(\X)+\eps (1+\|\X\|_2^2)^t$ is a sum of squares.
\end{thm}

\begin{rmk}
	Theorem~\ref{thm:sos-under-pert} holds for any $f\in \R[\X]$ with $f(\bm{0})\geq 0$.
\end{rmk}
Before giving a proof let us recall some basics on Gram matrices of sums of squares polynomials.
Let $\ms$ be the vector of all ordered (with respect to any fixed order) monomials in $n+1$ variables $X_0, X_1, \dots, X_n$ of degree $t$; it has $\Ns=\binom{n+t}{t}$ elements.
If $g$ is a real homogeneous polynomial $g$ of degree $2t$ in $n+1$ variables, we can write it as 
\begin{align}\label{eq:Gram}
	g(X_0, \X) = [m_t]^{\top} G \, [m_t],
\end{align}
where $G$ is a real symmetric matrix of size $N_t\times N_t$.
Such matrices $G$ form an affine subspace defined by the equation \eqref{eq:Gram}. 
Then, it holds that, $g$ is a sum of squares (of real homogeneous polynomials) if and only if \eqref{eq:Gram} holds for some positive semidefinite matrix $G$
 \cite[Thm.~2.4]{clr-real-sos-95}, \cite[Thm.~3.1]{powers-certif-book};
we call such a matrix \emph{a Gram matrix} of $g$. 
By slightly abusing notation and terminology, we will refer to \eqref{eq:Gram} as a Gram matrix representation of $g$, even when it is not a sum of squares (or even a nonnegative polynomial).
Furthermore, $g$ belongs to the interior of the cone of sums of squares of degree $2t$ in $n+1$ variables if and only if there exists a positive definite Gram matrix satisfying \eqref{eq:Gram}.


\begin{proof}[{Proof of Thm.~\ref{thm:sos-under-pert}}]
Let $f^h$ be the homogenization of $f$. Then the homogenization of $\fte$ (with $t\geq d$) is given as 
\begin{equation}
	\label{eq:fteh}		
\fte^h(X_0, \X) = X_0^{2(t-d)} \,f^h(X_0, \X) + \eps\,  (X_0^2 +\|\X\|_2^2)^t
\, \in \RR[X_0, \X]_{2t}. 
\end{equation}

The property of being a sum of squares is preserved under (de)homogenization.
Thus, it suffices to prove that $\fte^h$ is SOS for a suitable $t$.
We prove this by showing that the Gram matrix of $\fte^h$, for a sufficiently large $t$, is positive semidefinite. 
We construct Gram matrices of the two summands of $\fte^h$
	separately. For this, we order the degree $t$ monomials by 
first putting $X_0^t$, $X_0^{t-1} X_1,\dots, X_0^{t-1}X_n$.

\medskip
\noindent
\emph{The Gram matrix of $(X_0^2+\Vert \X\Vert_2^2)^t$}
\medskip

Consider first the second summand of $\fte^h$, that is, $(X_0^2+\Vert \X\Vert^2_2)^t=\sum_{\vert\balpha\vert=t} {t\choose \balpha} X_0^{2\alpha_0}X_1^{2\alpha_1}\dots X_n^{2\alpha_n}$.
It has a diagonal Gram matrix $A_t$ with the first diagonal entry being $1$. The matrix $A_t$ has the form 
\[
A_t=
\diag \Big\{ \binom{t}{\balpha} \Big\}_{ |\alpha | = t }= 
\begin{bmatrix}
  1 & 0         & 0 \\
  0 & A_{t,1}   & 0 \\
  0 & 0         & A_{t,2}
\end{bmatrix},
\]
where 
$A_{t,1} = \diag \Big\{ {t\choose \balpha}: t-d\leq \alpha_0 < t \Big\}$,
and
$A_{t,2}  = \diag \Big\{ {t\choose \balpha}: \alpha_0<t-d \Big\}$
for $\balpha = (\alpha_0, \alpha_1, \dots, \alpha_n)$.
We now bound smallest eigenvalues of $A_{t,1}$ and $A_{t,2}$. 
First, if $s=t-\alpha_0\in [d]$, for $t\ge 2d$, 
then
\begin{equation}
	\label{eq:lmin-At1}
	(A_{t,1})_{\alpha,\alpha}=\binom{t}{\balpha}\ge\binom{t}{s}\ge\binom{t}{1}=t\ 
\Rightarrow\ \lambda_{\min}(A_{t,1})\ge t .
\end{equation}

On the other hand, $A_{t,2}$ has some (diagonal) entries that correspond to  
the indices $t \bm{e}_i$ for $i \ge 1$, and,
using the monotonicity of the multinomial coefficients, we have  
$\lambda_{\min}(A_{t,2})=1$.

\medskip	
\noindent
\emph{The Gram matrix of the first summand}
\medskip

We consider a particular Gram matrix $F_d$ of $f^h(X_0, \X)=\md^\top F_d \,\md$
that is of the form 
\[
F_d 
=
\begin{bmatrix}
  f_0 & \vv^\top \\
  \vv   & \wt{F}_d
\end{bmatrix},
\]
where $f_0 = f(\bm{0}) = f^h(1,0,\dots, 0)$, 
\[ \vv^{\top}=\tfrac{1}{2}\left( f_{(2d-1)\bm{e}_0+ \bm{e}_1}, \dots, f_{(2d-1)\bm{e}_0+ \bm{e}_n}, 0,\dots, 0\right)
	= 
\tfrac{1}{2} \left(\partial_1 f(\bm{0}),\dots, \partial_n f(\bm{0}), 0,\dots, 0\right).
\]
and such that elements of the submatrix $\wt{F}_d$ depend linearly on the coefficients of $f^h$.
The fact that we can obtain the degree $2d$ monomials $(X_0, \X)^{d \bm{e}_0+ \bm{\alpha}}$, 
with $\bm{\alpha} \neq (d-1) \bm{e}_0+ \bm{e}_i$ (with some $i=0,1,\dots,n$),
in more than one way (as a product of two degree $d$ monomials) 
justifies that $F_d$ has the required form.

We set $F_t$ to be a $N_t\times N_t$ symmetric matrix whose entries are 
$(F_t)_{(t-d)\bm{e}_0+ \bm{\alpha}, (t-d) \bm{e}_0+ \bm{\beta}}  
= (F_d)_{\bm{\alpha}, \bm{\beta}}$,
with  $\abs{\bm{\alpha}} = \abs{\bm{\beta}} =d$.
Then,  
\[
X_0^{2(t-d)} f^h(X_0, \X) = 
X_0^{2(t-d)} \md^{\top}\, F_d\, \md = 
\mt^{\top}\, F_t\, \mt, 
\]
that is, $F_t$ is a Gram matrix representation of $X_0^{2(t-d)} f^h(X_0, \X)$.
 Overall, the $N_t\times N_t$ symmetric matrix $F_t$ has a block-diagonal structure
\begin{equation}
    F_t = 
\begin{pmatrix}
    f_0 & \vv^{\top} & {\large 0}\\
    \vv & \wt{F}_d & {\large 0} \\
    {\large 0} & {\large 0} & {\large 0}
\end{pmatrix},
\end{equation}
where the entries of the submatrix $F_d$ of size $N_d\times N_d$ are labeled by $(t-d)\, \bm{e}_0+ \bm{\alpha}$ and $(t-d)\, \bm{e}_0+ \bm{\beta}$.


\medskip

\medskip
\noindent
\emph{The Gram matrix of $\feps^h$ and its eigenvalues}
\medskip	

Consider $G_t:=F_t+\eps A_t$.
Then $\fte^h = \mt^\top G_t \, \mt$, that is, $G_t$ is a Gram matrix for $\fte^h$. 
We have
\[
G_t
=
\begin{bmatrix}
  f_0 + \eps & \vv^\top & 0 \\[5pt]
  \vv & \wt{F}_d + \eps A_{t,1} & 0 \\[5pt]
  0 & 0 & \eps A_{t,2}
\end{bmatrix}.
\]
For the lower (bottom right) block, it holds 
 $\eps \, A_{t,2} \succeq \eps I$. In particular, it is always positive definite 
 and $\lambda_{\min}(\eps A_{t,2})=\eps$.
So, $G_t$ is positive semi-definite ($G\succeq 0$) if and only if so is its submatrix
\[
\begin{bmatrix}
   f_0 + \eps & \vv^\top   \\[5pt]
  \vv & \wt{F}_d + \eps A_{t,1}\\[5pt]
\end{bmatrix}.
\]
As the constant term of a nonnegative polynomial is nonnegative, we have $f_0+\eps>0$.
Thus, $G_t \succeq 0$ if and only if the following Schur complement is positive semi-definite \cite[Thm. 7.7.7]{HornJohnson}:
\[
\wt{F}_d + \eps \, A_{t,1} - \frac{\vv \vv^\top}{f_0+\eps}  \succeq 0.
\]
To ensure that the smallest eigenvalue of this matrix is nonnegative we use $\lambda_{\min}(M_1 + M_2) \geq \lambda_{\min}(M_1) + \lambda_{\min}(M_2)$.
Therefore, it is sufficient to prove that
\[ 
t\,\varepsilon +\lambda_{\min}(\wt{F}_d) +\lambda_{\min}\left(- \frac{\vv \vv^\top}{f_0+\eps}\right)\geq 0,
\]
where we use $\lambda_{\min}(A_{t,1})\ge t$ shown in \eqref{eq:lmin-At1}.
Now, using the inequality $\lambda_{\min}(M) \geq - \ntwo{M}$ and $\Vert \vv \vv^\top\Vert_2 = \Vert \vv\Vert_2^2$, it is enough to ensure that 
\begin{align}\label{eq:local1}
 t \, \eps - \ntwo{\wt{F}_d}  - \frac{ \ntwo{\vv}^2}{f_0+\eps} \geq 0.
\end{align}
Using the properties of the operator norm, we have that 
\[
\ntwo{\wt{F}_d} \leq \ntwo{F_d} \leq \sqrt{\ninf{F_d} \, \none{F_d}} = \none{F_d}
\leq \none{f^h} = \none{f}  \enspace. 
\]
This, together with $\Vert \vv\Vert_2=\frac{1}{2}\Vert \nabla f(\bm{0})\Vert_2\leq \Vert \nabla f(\bm{0})\Vert_2$, implies that the bound \eqref{eq:local1} holds if 
\[
t \geq \frac{1}{\varepsilon}\left(\Vert f\Vert_1+\frac{\ntwo{\nabla{f}(\bm{0})}^2 }{f_0+ \eps}\right).
\]
\end{proof}

\begin{rmk}
	Let $f \in \ZZ[\X]$ be any positive polynomial in $n$ variables, of degree $2d$, and $H(f) \leq 2^{\tau}$. 
	Also, let $\eps = 2^{-L}$, for a positive integer $L$.
	Then, Thm.~\ref{thm:sos-under-pert} implies that for any $t$, such that
	\[
		t \geq \binom{n + 2d}{n} \, 2^{L + \tau + 1},
	\]
	the polynomial $f(\X) + \eps (1 + \ntwo{\X}^2)^t$ is a sum of squares.
\end{rmk}

\subsection{Comparison with Lasserre's perturbations}\label{sub:comparison}

The bound on $t$ of Theorem~\ref{thm:sos-under-pert} seems to be
quite non-optimal for a concrete polynomial. For example, the perturbation $M+\varepsilon (1+X_1^2+X_2^2)^t$ of the Motzkin polynomial  $M=X_1^2X_2^4+X_1^4X_2^2+1-3X_1^2X_2^2$ is a sum of squares for all $t>\max\{3,6/\varepsilon\}$ by Theorem~\ref{thm:sos-under-pert}.
Using JuMP/MathOptInterface \cite{JuMP, MOI} in Julia and MOSEK \cite{mosek2024} for solving an SDP, we find that the smallest eigenvalue of $\tilde F_d$ among all Gram matrices as in the proof of Theorem~\ref{thm:sos-under-pert} is $\lambda_{\min}(\tilde F_d)\gtrsim -0.23$ which implies that $M+\varepsilon (1+X_1^2+X_2^2)^t$ is a sum of squares already for $t\geq \max\{3,0.23/\varepsilon\}$.
Allowing all (that is, not necessarily structured as above) Gram matrices reveals that $M+\varepsilon (1+X_1^2+X_2^2)^t$ is a sum of squares for $\varepsilon=0.0002$, $t=4$ and $\varepsilon=0.0000027$, $t=5$.

As we already mentioned above, Lasserre proved in \cite{l-sos-approx-06} that for a nonnegative $f\in \R[\X]$ and any $\eps>0$, for all $t\in \mathbb{N}$ sufficiently large, the polynomial $f+\eps \theta_t$ is a sum of squares,
 where $\theta_t= \sum_{k=0}^t\sum_{i=1}^n \frac{X_i^{2k}
}{k!}$. Let us look at the expansion of the polynomial appearing in the perturbation 
\eqref{eq:fte},
$$ h_t\ :=\ (1+X_1^2+\dots+X_n^2)^{t}\ =\ \sum_{\vert \balpha\vert\leq t} \frac{t!}{(t-\vert \balpha\vert)!\balpha!} \,\X^{2\balpha},$$
which is a sum of squares of monomials. Looking only at the terms with $\alpha=ke_i$ for $k=0,\dots, t$ and $i=1,\dots, n$, we obtain
$$ \sum_{k=0}^t \sum_{i=1}^n \frac{t!}{(t-k)!k!}\, X_i^{2k}\ =\ \theta_t+\sum_{k=0}^t \sum_{i=1}^n \left(\frac{t!}{(t-k)!}-1\right)\frac{X_i^{2k}}{k!},$$ that is,
a sum of $\theta_t$ and a sum of squares of some monomials. The remaining terms in $h_t$ are also squared monomials. As a consequence, for all $t\in \mathbb{N}$ for which Lasserre's perturbation $f+\varepsilon \theta_t$ is a sum of squares we also have that $f+\varepsilon h_t$ is a sum of squares.
However, it is not fair to compare these two perturbations as they have different sizes. For example, the $1$-norm of coefficients of the perturbations are $\Vert \theta_t\Vert_1= ns_t$, where $s_t:=\sum_{k=0}^t \frac{1}{k!}\leq e$ is bounded by a constant independent of $t$, and $\Vert h_t\Vert_1=(n+1)^t$ which is exponential in $t$.
We compare the normalized perturbations $f+ \varepsilon \tilde \theta_t$ with $\tilde\theta_t:=\frac{\theta_t}{n s_t}$ and $f+\varepsilon \tilde h_t$ with $\tilde h_t:=\frac{h_t}{(n+1)^t}$. For each of the two models we computed the smallest $\varepsilon>0$ for which a normalized perturbation of the Motzkin polynomial $f=M=X_1^2X_2^4+X_1^4X_2^2+1-3X_1^2X_2^2$ is a sum of squares, below we display the results rounded up to two significant digits. 

\begin{equation}\begin{tabular}{|c|c|c|c|c|c|}\hline 
& $t=3$ & $t=4$ & $t=5$ & $t=6$  \\ \hline
  $\phantom{\Big|}M+\varepsilon \tilde \theta_t$ & $0.14$ & $0.025$ & $0.0023$ & $0.00011$ \\  \hline 
 $\phantom{\Big||}M+\varepsilon \tilde h_t$ & $0.13$ & $0.016$ & $0.00063$ & $0.000011$  \\ \hline
\end{tabular}\end{equation}

 Lasserre showed \cite{l-norm1-approx-10} that, given a nonnegative polynomial $f\in \R[\X]$, a sum of squares polynomial of degree at most $2t\geq \deg f$ that is closest to $f$ with respect to the $1$-norm of coefficients is obtained as $f+\lambda_0^*+\sum_{i=1}^n \lambda_i^* X_i^{2t}$ for some $\lambda_0^*,\lambda_1^*,\dots, \lambda_n^*\in \R$. The perturbation $f+\varepsilon \theta_t$ from 
 \cite{l-sos-approx-06} has a similar form to this one. In the following table we compare the $1$-norm of errors of three SOS perturbations of a Motzkin-like polynomial $\tilde M=X_1^2X_2^4+X_1^4X_2^2+1-3X_1^2X_2^2$ considered in \cite[Table $1$]{l-norm1-approx-10}.

\begin{equation}\begin{tabular}{|c|c|c|c|c|}\hline 
& $t=3$ & $t=4$ & $t=5$  \\ \hline
$\tilde M+\lambda_0^*+\lambda_1^*x_1^{2t}+\lambda_2^*x_2^{2t}$ & $0.016$ & $0.0021$ & $0.000087$ \\ \hline
  $\phantom{\Big|}\tilde M+\varepsilon \tilde \theta_t$ & $0.037$ & $0.016$ &  $0.0053$ \\  \hline 
 $\phantom{\Big|}\tilde M+\varepsilon \tilde h_t$ & $0.054$ & $0.023$ & $0.0038$  \\ \hline
\end{tabular}\end{equation}

A universal perturbation valid for all (or almost all) nonnegative polynomials of a given degree might not exist.
Even though the class of perturbations considered in  \cite{l-norm1-approx-10} is rather simple and contains optimal ones (in $1$-norm), for each given $f\in \R[\X]$ one needs to solve a semi-definite program.
The above two tables show that the normalized perturbations $f+\varepsilon \tilde h_t$ and $f+\varepsilon \tilde\theta_t$ are in general incomparable. For the Motzkin polynomial $M$ at least for $t=3, 4, 5$ and $6$, ``moving'' from $M$ in the direction of $\tilde h_t$ results in a sum of squares polynomial a bit faster than moving along $\tilde \theta_t$. On the other hand, the polynomial $\tilde M$ is closer in $1$-norm to sums of squares of the form $\tilde M+\varepsilon  \theta_t$ than to sums of squares of the form $\tilde M+\varepsilon h_t$ for $t=3$ and $t=4$, but for $t=5$ the proximity relation reverses. We plan to carry out a careful comparison of these two natural perturbations in a future work.

\section*{Acknowledgements}
This research benefited from the support of the 
FMJH Program Gaspard Monge for optimization and operations 
research and their interactions with data science through the PGMO grant SOAP.
MB is supported by the grant
ANR JCJC PeACE (ANR-25-CE48-3760).
KK is supported by the grant ANR ``Chaire de Professeur Junior Mathématiques, Applications des Mathématiques, Géométrie Computationnelle GeoComp''.
ET is supported by the grant ANR PRC ZADyG (ANR-25-CE48-7058). 
CZ is supported by the Chinese Scholarship Council.

\printbibliography

\end{document}